\documentclass[aps,prl,floatfix,twocolumn,showpacs,nofitinbib,hyperref=pdftex,citeautoscript,superscriptaddress]{revtex4-1}
\usepackage{epsfig}
\usepackage{amsmath}
\usepackage{amssymb}
\usepackage{amsfonts}
\usepackage{ dsfont }
\usepackage{ comment,color }
\usepackage{lipsum}
\usepackage{hyperref,mathtools}
\usepackage{makecell}
\usepackage{array}
\usepackage{placeins}

\hypersetup{colorlinks=true,linkcolor=blue,anchorcolor=blue,citecolor=blue,filecolor=blue,urlcolor=blue,bookmarksnumbered=true,pdfview=FitB}

\newcommand{\beqa}{\begin{eqnarray}}
\newcommand{\eeqa}{\end{eqnarray}}

\begin{document}

\hsize\textwidth\columnwidth\hsize\csname@twocolumnfalse\endcsname

\title{Supercurrent parity-meter in a nanowire Cooper-pair transistor}
\affiliation{QuTech and Kavli Institute of NanoScience, Delft University of Technology, 2600 GA Delft, The Netherlands}
\affiliation{Department of Physics, Massachusetts Institute of Technology, 77 Massachusetts Avenue, Cambridge, Massachusetts 02139 USA}
\affiliation{Beijing National Laboratory for Condensed Matter Physics, Institute of Physics, Chinese Academy of Sciences, Beijing 100190, China}
\affiliation{Department of Applied Physics, Eindhoven University of Technology, 5600 MB Eindhoven, The Netherlands}
\affiliation{California NanoSystems Institute, University of California Santa Barbara, Santa Barbara, California 93106, USA}
\affiliation{Current address: Department of Physics and Astronomy, University of Tennessee, Knoxville, Tennessee 37996, USA}
\affiliation{Electrical and Computer Engineering, University of California Santa Barbara, Santa Barbara, California 93106, USA}
\affiliation{Materials Department, University of California Santa Barbara, Santa Barbara, California 93106, USA}
\affiliation{Microsoft Station Q Delft, 2600 GA Delft, The Netherlands}
\affiliation{These authors contributed equally}

\author{Ji-Yin Wang}
\address{QuTech and Kavli Institute of NanoScience, Delft University of Technology, 2600 GA Delft, The Netherlands}
\address{These authors contributed equally}

\author{Constantin Schrade}
\address{Department of Physics, Massachusetts Institute of Technology, 77 Massachusetts Avenue, Cambridge, Massachusetts 02139 USA}
\address{These authors contributed equally}

\author{Vukan Levajac}
\author{David van Driel}
\author{Kongyi Li}
\address{QuTech and Kavli Institute of NanoScience, Delft University of Technology, 2600 GA Delft, The Netherlands}

\author{Sasa Gazibegovic}
\author{Ghada Badawy}
\author{Roy L.M. Op het Veld}
\address{Department of Applied Physics, Eindhoven University of Technology, 5600 MB Eindhoven, The Netherlands}

\author{Joon Sue Lee}
\affiliation{California NanoSystems Institute, University of California Santa Barbara, Santa Barbara, California 93106, USA}
\affiliation{Current address: Department of Physics and Astronomy, University of Tennessee, Knoxville, Tennessee 37996, USA}

\author{Mihir Pendharkar}
\address{Electrical and Computer Engineering, University of California Santa Barbara, Santa Barbara, California 93106, USA}

\author{Connor P. Dempsey}
\address{Electrical and Computer Engineering, University of California Santa Barbara, Santa Barbara, California 93106, USA}

\author{Chris J. Palmstrøm}
\address{California NanoSystems Institute, University of California Santa Barbara, Santa Barbara, California 93106, USA}
\address{Electrical and Computer Engineering, University of California Santa Barbara, Santa Barbara, California 93106, USA}
\address{Materials Department, University of California Santa Barbara, Santa Barbara, California 93106, USA}

\author{Erik P.A.M. Bakkers}
\address{Department of Applied Physics, Eindhoven University of Technology, 5600 MB Eindhoven, The Netherlands}

\author{Liang Fu}
\address{Department of Physics, Massachusetts Institute of Technology, 77 Massachusetts Avenue, Cambridge, Massachusetts 02139 USA}

\author{Leo P. Kouwenhoven}
\address{QuTech and Kavli Institute of NanoScience, Delft University of Technology, 2600 GA Delft, The Netherlands}
\address{Microsoft Station Q Delft, 2600 GA Delft, The Netherlands}

\author{Jie Shen}
\email{shenjie@iphy.ac.cn}
\address{QuTech and Kavli Institute of NanoScience, Delft University of Technology, 2600 GA Delft, The Netherlands}
\address{Beijing National Laboratory for Condensed Matter Physics, Institute of Physics, Chinese Academy of Sciences, Beijing 100190, China}

\date{\today}

\vskip1.5truecm
\begin{abstract}
We study a Cooper-pair transistor realized by two Josephson weak links that enclose a superconducting island in an InSb-Al hybrid nanowire. When the nanowire is subject to a magnetic field, isolated subgap levels arise in the superconducting island and, due to the Coulomb blockade, mediate a supercurrent by coherent co-tunneling of Cooper pairs. We show that the supercurrent resulting from such co-tunneling events exhibits, for low to moderate magnetic fields, a phase offset that discriminates even and odd charge ground states on the superconducting island. Notably, this phase offset persists when a subgap state approaches zero energy and, based on theoretical considerations, permits parity measurements of subgap states by supercurrent interferometry. Such supercurrent parity measurements could, in a new series of experiments, provide an alternative approach for manipulating and protecting quantum information stored in the isolated subgap levels of superconducting islands. 
\end{abstract}


\maketitle
When two superconducting (SC) leads couple via a Coulomb-blockaded quantum dot (QD), the isolated energy levels on the dot mediate a supercurrent by coherent co-tunneling of Cooper pairs \cite{bib:Spivak1991}. For the particular case of a single-level QD, a control knob for the direction of the supercurrent is given by the charge parity of dot electrons \cite{bib:Spivak1991}. Such a parity-controlled supercurrent has been observed in a nanowire (NW) QD Josephson junction (JJ) \cite{bib:vanDam2006,bib:Razmadze2020}. 
It is described by the Josephson relation, $I=(-1)^{n_{0}}I_{c}\sin(\varphi)$, where $I_{c}$ is the critical current, $\varphi$ is the SC phase difference, and $n_{0}$ is the number of dot electrons. 
In general, the Josephson relation can also acquire a phase offset, $\varphi\rightarrow\varphi+\varphi_0$ with $\varphi_{0}\neq0,\pi$, when time-reversal and mirror symmetry are broken \cite{bib:Zazunov2009}. This breaking occurs, for example, if a spin-orbit coupled QD is subject to a magnetic field \cite{bib:Zazunov2009,bib:Brunetti2013,bib:Szombati2016,bib:Schrade2017}. 

A different possibility of coupling two SC leads is via an intermediate SC island with finite charging energy; a `Cooper-pair transistor' (CPT) \cite{bib:Geerligs1990,bib:Tuominen1992,bib:Fulton1989,bib:Woerkom2015,bib:Veen2018,bib:Proutski2019,bib:Schrade20184}. Unlike in the QD JJ, the SC island carries, within its parity lifetime, an even number of electrons in the ground state, as signified by a charging energy that is a $2e$-periodic function of the island gate charge (\textit{e}, elementary charge) \cite{bib:Tuominen1992,bib:Woerkom2015,bib:Veen2018}. In particular, since the odd charge states are energetically unfavorable for a conventional CPT, the 
Josephson relation is not expected to exhibit a parity-controlled phase offset.

\begin{figure}[!t] \centering
\includegraphics[width=0.98\linewidth] {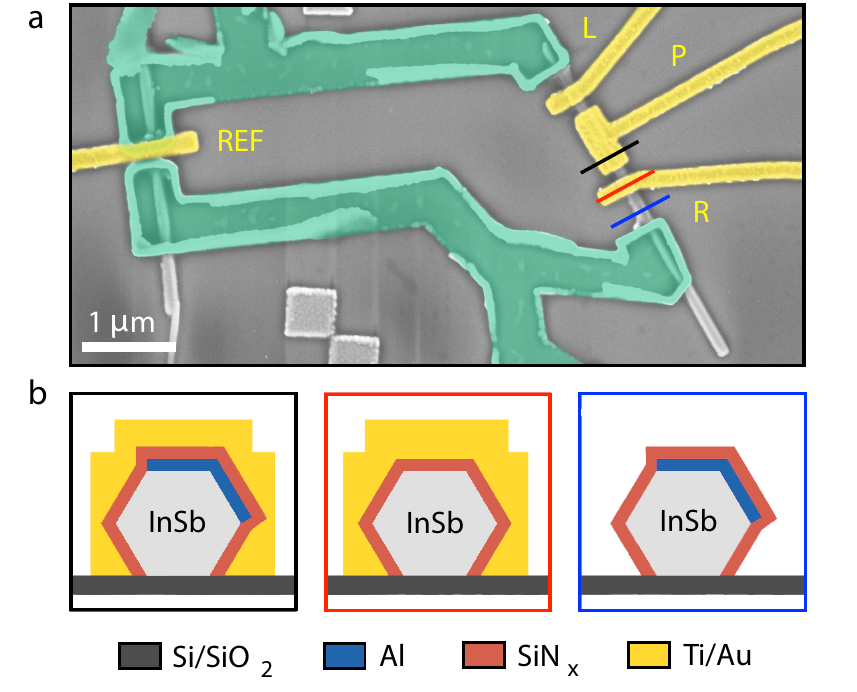}
\caption{(Color online)
\textbf{a}, False-color micrograph of the measured NbTiN (green) SQUID device comprising an InSb-Al NW CPT in the right arm and an InSb nanowire reference junction in the left arm. Top gates (L,\,R,\,REF) define tunable JJs, and a plunger gate (P) controls the electron number on the SC island of the nanowire CPT. \textbf{b}, Cross-sections along the lines shown in (a). Gates are made of Ti/Au ($10/120$\,nm) and are separated from the InSb-Al nanowire by $30$\,nm sputtered SiN$_x$. The substrate, p-doped Si covered with $285$\,nm SiO$_2$, acts as a global back gate.
}\label{fig:1}
\end{figure}

Recently, a CPT has been realized with an Indium Arsenide-Aluminium (Al) hybrid NW \cite{bib:Veen2018,bib:Proutski2019}. In this case, upon increasing a magnetic field parallel to the NW, a transition from a $2e$-periodic switching current to a switching current with even-odd pattern has been observed \cite{bib:Veen2018}. The interpretation is that a low-energy subgap state arises in the SC island, and, depending on its occupancy, the charge ground state carries an even or an odd number of electrons. An open question is if the Josephson relation of a NW CPT exhibits \textit{in the presence of subgap states} a parity-controlled phase offset? 

Here, we address this question with a NW CPT integrated in a superconducting quantum interference device (SQUID). We investigate the previously described situation when the NW CPT is subject to a parallel magnetic field so that subgap levels arise in the SC island and mediate a supercurrent by coherent co-tunneling of Cooper pairs. We show that supercurrent resulting from Cooper pair co-tunneling exhibits a phase offset, which distinguishes even and odd charge ground states on the SC island. This phase offset persists when a subgap state approaches zero energy and, based on theoretical considerations, enables \textit{parity readout} of low-energy subgap states. Such supercurrent parity readout could provide a new approach for the manipulation \cite{bib:Beenakker2004,bib:Mao2004,bib:Engel2005,bib:Lalumiere2010,bib:Ionicioiu2007,bib:Pfaff2013} and protection \cite{bib:Andersen2019,bib:Bultink2020} of quantum information stored in the isolated subgap levels of SC islands \cite{bib:Aasen2016,bib:Plugge2017,bib:Karzig2017,bib:Schrade20182,bib:Schrade20183}.
 
The device geometry of our experiment is shown in Fig.~\ref{fig:1}. For realizing the CPT, we use a shadow-grown \cite{bib:Gazibegovich2017} Al SC island on an Indium Antimonide (InSb) NW, which couples to two SC Al leads via gate-tunable tunneling barriers. A plunger gate is used for controlling the electron number on the SC island. As we intend to study the full Josephson relation of the NW CPT, we integrate our setup in a SQUID loop made of niobium-titanium nitride (NbTiN) and a second InSb NW reference junction. The tunnel coupling of the reference junction is adjustable by a local gate electrode.

\begin{figure}[!t] \centering
\includegraphics[width=1\linewidth] {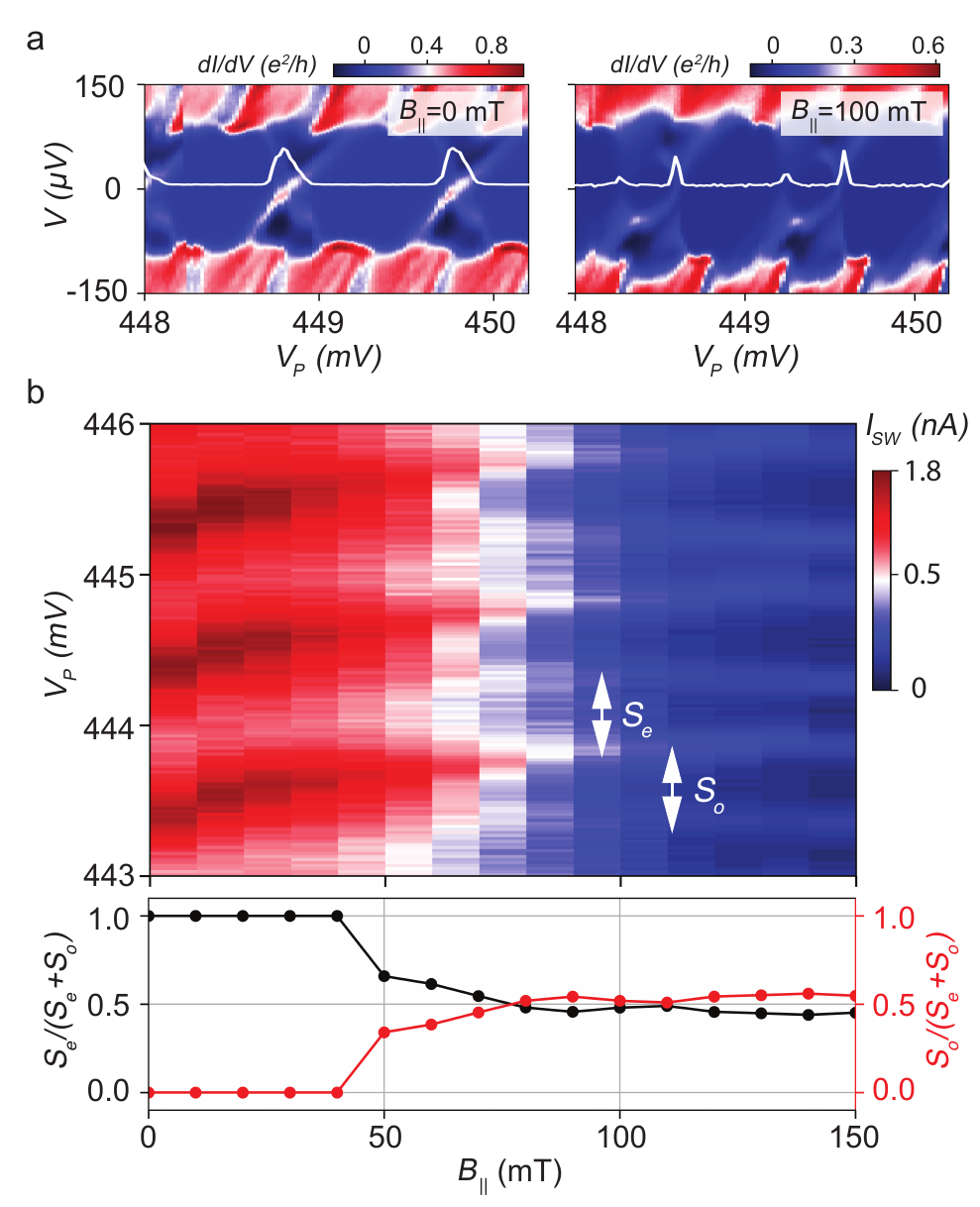}
\caption{(Color online)
\textbf{a}, Differential conductance, $\text{d}I/\text{d}V$, versus source-drain voltage $V$ and plunger gate voltage $V_{P}$. At zero parallel magnetic field, the differential conductance shows a Coulomb diamond pattern with a $2e$-periodicity. At $B_{\parallel}=100\,$mT, the $2e$-periodicity of the Coulomb diamonds lifts due to the appearance of an odd-parity charge ground state on the SC island. Inset curves show the differential conductance at zero bias. \textbf{b}, Top panel: Switching current, $I_{sw}$, versus parallel magnetic field $B_{\parallel}$ and plunger gate voltage $V_{P}$. Bottom panel: Magnetic field dependence of the normalized even and odd peak spacings, $S_{e}/(S_{e}+S_{o})$ and $S_{o}/(S_{e}+S_{o})$, showing a transition from a $2e$-periodicity to an even-odd pattern. Such a transition in the peak spacing is consistent with a state detaching from the SC island's quasiparticle continuum and forming an isolated subgap level \cite{bib:Veen2018}.
}\label{fig:2}
\end{figure}

Initially, we pinch off the reference junction and characterize the NW CPT by measuring the differential conductance $\mathrm{d}I/\mathrm{d}V$ versus the source-drain voltage $V$ and the plunger gate voltage $V_{P}$. Our results are shown in  Fig.~\ref{fig:2}a for zero and finite parallel magnetic fields $B_{\parallel}$.

At zero magnetic field, we observe a pattern of Coulomb diamonds with sharp edges due to the weak island-lead coupling. Besides the Coulomb diamonds, which signify the importance of charging effects on the SC island, we also measure conductance peaks at zero-bias around the SC island's charge degeneracy points (insert curve in Fig.~\ref{fig:2}a). These peaks exhibit a $2e$-periodicity and indicate that, in addition to charging effects, the tunneling of Cooper pairs is also relevant in the presented transport regime. Furthermore, above a onset voltage $V_{onset}$, a $1e$-periodic modulation of the differential conductance appears, which marks the onset of quasiparticle transport. 

At finite magnetic fields, the aforementioned onset voltage for quasiparticle transport persists. However, below the onset voltage, the Coulomb diamonds split, resulting in an even-odd pattern. We attribute the appearance of this even-odd pattern to low-energy subgap states that form on the SC island. More specifically, the magnetic field induces a Zeeman splitting of spinful, odd-parity states and, thereby, reduces the minimum single-particle excitation energy in the NW CPT. As a result, odd-parity states that partially reside in the InSb nanowire can detach from the quasiparticle continuum and, because of their enhanced effective $g$-factor in comparison to the Al shell, form isolated levels below the SC gap \cite{bib:Veen2018,bib:Vaitiekenas2018}.

Next, we investigate the subgap levels on the SC island in more detail. We lower the island-lead tunneling barriers and, with the reference junction still pinched off, measure the switching current $I_{sw}$ as a function of the parallel magnetic field $B_{\parallel}$ and plunger gate voltage $V_{P}$. Our results are depicted in  Fig.~\ref{fig:2}b. At zero magnetic field, the switching current exhibits a $2e$-periodic peak spacing implying that the SC island always carries an even number of electrons in its charge ground state (see also Fig. S1a in supplementary material). The situation changes upon applying a parallel magnetic field. The magnetic field induces a splitting of the $2e$-periodic peaks, and, as a result, the switching current exhibits a peak-spacing with an even-odd pattern (see also Fig. S1b in supplementary material). Similar to the differential conductance, we attribute the appearance of this even-odd pattern to charge ground states with even and odd fermion parity on the SC island. Moreover, the extracted peak spacing oscillates as a function of applied magnetic field (bottom panel in fig.2b), as well as the plunger gate voltage, indicating either the anticrossing or the crossing of the lowest-energy subgap state with a second subgap state at higher energy\cite{bib:Jie2018,bib:Jie2020}.

\begin{figure*}[!t]
\centering
\includegraphics[width=\textwidth]{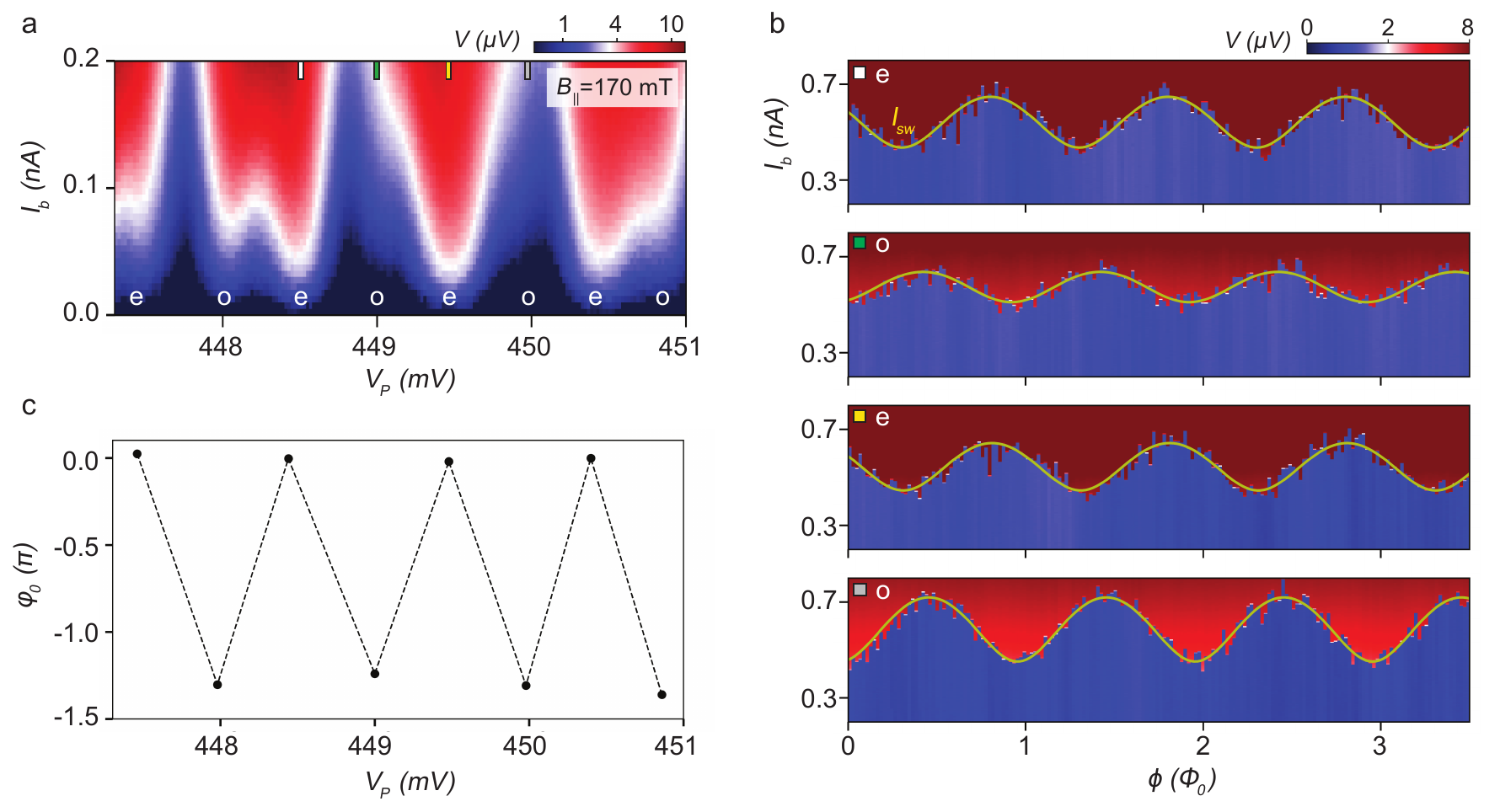}
\caption{(Color online) \textbf{a}, Voltage drop $V$ across the NW CPT versus applied bias current $I_{b}$ and plunger gate voltage $V_{P}$, showing an even-odd pattern consistent with the appearance of low-energy subgap states at a parallel magnetic field  $B_{\parallel}=170\,$mT. \textbf{b}, Voltage drop $V$ as a function of the applied bias current  $I_{b}$ and the flux $\Phi$ that pierces through the SQUID loop for the plunger gate voltages $V_{P}$ marked in (a). The fitted switching current $I_{sw}$ (yellow) displays a phase offset $\varphi_{0}$ that discriminates the even and odd charge parity sectors of the SC island. \textbf{c}, Phase offset $\varphi_{0}$ (relative to the even Coulomb valleys) versus plunger gate voltage $V_{P}$. Dashed lines are a guide to the eye. In the range of plunger gate voltages shown here, the phase offset is insensitive. 
}\label{fig:3}
\end{figure*}

We now open the reference junction and measure the NW CPT's full Josephson relation in the presence of low-energy subgap states. For the results presented here, we focus on the magnetic field strength $B_{\parallel}=170$\,mT, and adopt a highly-asymmetric SQUID configuration so that the phase drop occurs primarily across the NW CPT. Under these conditions, we apply a bias current $I_{b}$ and measure the voltage drop $V$ across the SQUID as a function of the plunger gate voltage $V_{P}$ and the flux $\Phi$ piercing through the SC loop. Fig.~\ref{fig:3} shows our measurement data, which we will now discuss in more detail:

The main result of our measurements is that the Josephson relation of the NW CPT exhibits a substantial relative phase offset $\varphi_0$ between Coulomb valleys of opposite charge parity. To determine this phase offset for the Coulomb valleys marked in  Fig.~\ref{fig:3}a, we fit the switching current $I_{sw}$ as a function of the flux $\Phi$. The fitted curves, shown in  Fig.~\ref{fig:3}b, allow us to extract $\varphi_{0}\sim-1.24\pi$ and $\varphi_{0}\sim-1.31\pi$ for the first and second pair of Coulomb valleys, respectively. For the remaining pairs, we find similar values for the phase offset, as summarized in  Fig.~\ref{fig:3}c. Notably, the leftmost pair of data points in  Fig.~\ref{fig:3}c shows that 
phase offset persists when the Coulomb peaks are approximately $1e$-spaced (see detailed analysis in Fig. S2 in supplementary materials). Therefore, the phase offset facilitates charge parity readout even if a subgap state approaches zero energy, and the behavior of the SC island's addition energy is \textit{identical} for opposite charge parity sectors.

\begin{figure}[!t] \centering
\includegraphics[width=1\linewidth] {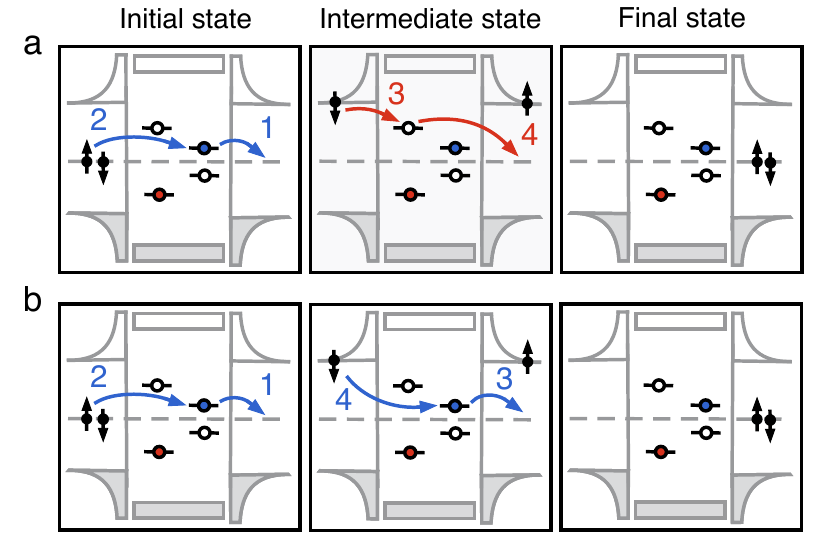}
\caption{(Color online)
\textbf{a}, A typical sequence of intermediate states in which a Cooper pair tunnels between the SC leads (left, right) via the two lowest isolated subgap levels $a,b$ in the intermediate SC island (center). Such a sequence yields a contribution to the supercurrent proportional to the joint parity of the two subgap levels. In the illustration, numbers indicate the sequence of tunneling events, and solid/empty dots represent filled/empty subgap levels.
\textbf{b}, A typical sequence of intermediate states that involves Cooper pair transport via a single subgap level yielding no parity-dependent prefactor.
}\label{fig:4}
\end{figure}

Motivated by our measurements, we now discuss a mechanism for a parity-dependent phase offset. We introduce a model for the NW CPT, which comprises a mesoscopic SC island coupled to a pair of $s$-wave SC leads. In our model, we focus on the two lowest isolated subgap levels in the SC island, $\pm E_{a}$ and $\pm E_{b}$,indicated by the peak spacing oscillation as a function of magnetic field and plunger gate in ~\ref{fig:2}b. These two levels form an effective spin degree of freedom if the Coulomb blockade fixes the SC island's total fermion parity. In particular, the effective spin mediates a supercurrent, similar to the electron spin in a QD JJ, through the co-tunneling of Cooper pairs. We distinguish between two types of co-tunneling sequences: 

(1) In the first type of sequence, shown in Fig.~\ref{fig:4}a, the Cooper pair splits so that one electron tunnels via $\pm E_{a}$ while the other electron tunnels via $\pm E_{b}$. 
For such a two-level sequence, the corresponding supercurrent contribution acquires a prefactor given by the SC island charge parity, $(-1)^{n_{0}}$. This parity prefactor is analogous to the parity prefactor appearing in the Josephson relation of a QD JJ, where Cooper pairs tunnel via two dot levels with opposite spin polarization \cite{bib:Spivak1991}. 

(2) In the second type of sequence, shown in Fig.~\ref{fig:4}b, both Cooper pair electrons tunnel via either $\pm E_{a}$ or $\pm E_{b}$. For such a single-level sequence, each of the two electrons contributes a prefactor given by the parity of $\pm E_{a}$ or $\pm E_{b}$. In particular, since the same parity prefactor appears twice in the sequence, it squares to one. Consequently, in the single-level supercurrent contribution a parity prefactor is absent.

If we collect all co-tunneling sequences of the first and second type, we obtain the Josephson relation (see details in supplementary material section 4),
\begin{equation}
\label{Eq1}
I = (-1)^{n_{0}}I_{ab}\sin(\varphi+\varphi_{ab})
+
\sum_{\ell=a,b}
I_{\ell}\sin(\varphi+\varphi_{\ell}).
\end{equation}
Here, $I_{ab}$ and $I_{\ell}$ are amplitudes, which are $1e$-periodic in the gate charge if the lowest subgap level is at zero energy. Furthermore, $\varphi_{ab},\varphi_{\ell}$ are phase offsets due to the simultaneous breaking of time-reversal and mirror symmetry. In our setup, time-reversal symmetry breaking arises from the magnetic field, inducing different tunneling probabilities for spin-up and spin-down electrons. The breaking of mirror symmetry results from asymmetric tunneling contacts so that Cooper pairs tunneling across the left/right JJ acquire different tunneling phases. 

At this point, it is worth highlighting two differences between the NW CPT and a QD JJ: First, the island which mediates the Josephson current is in a SC state, not a
normal state as for a QD JJ. Consequently, not only conventional tunneling events can occur, but also anomalous tunneling events in which an electron is created/destroyed on both the SC island and the leads. Second, for a QD JJ, the wavefuntions on the dot are highly localized which justifies a point-like tunneling contact. In comparison, for a NW CPT, the subgap level wavefunctions can be extended, which induces longer-range island-lead tunnel couplings. In particular, such longer-range couplings can break the mirror symmetry, due to the combined effect of spin-orbit coupling and magnetic field in the tunneling region, and lead to additional contributions to $\varphi_{ab},\varphi_{\ell}$ .

Returning to Eq.~\eqref{Eq1}, the total phase offset is $\varphi_{0}\equiv\text{arg}[-I_{ab}e^{i\varphi_{ab}}+\sum_{\ell}I_{\ell}e^{i\varphi_{\ell}}]-\text{arg}[I_{ab}e^{i\varphi_{ab}}+\sum_{\ell}I_{\ell}e^{i\varphi_{\ell}}]$. In this expression, the parity prefactor flips upon tuning the gate charge of the SC island between different charge parity sectors. As a result of these parity-flips, the phase offset does \textit{not} exhibit a $1e$-periodicity in the gate charge even if one of the subgap states is at zero energy. Instead, if $I_{ab}\neq0$, $\varphi_0$ is always $2e$-periodic and permits the measurement of the parity of the lowest subgap level. To practically enable such parity measurements, the two-level contribution should be sizable, $I_{ab}\gg I_{\ell}$. Also, to avoid thermal excitations, the temperature $T$ should be small compared to the level separation $|E_{a}-E_{b}|$. Interestingly though, if $|E_{a}-E_{b}|\gtrsim T$, the parity prefactor measures the joint parity of $\pm E_{a}$ and $\pm E_{b}$. Such joint parity measurements could be leveraged for 
entangling qubits stored in the subgap levels of SC islands \cite{bib:Beenakker2004,bib:Mao2004,bib:Engel2005,bib:Lalumiere2010,bib:Ionicioiu2007,bib:Pfaff2013}.

\begin{figure}[!t] \centering
\includegraphics[width=1\linewidth] {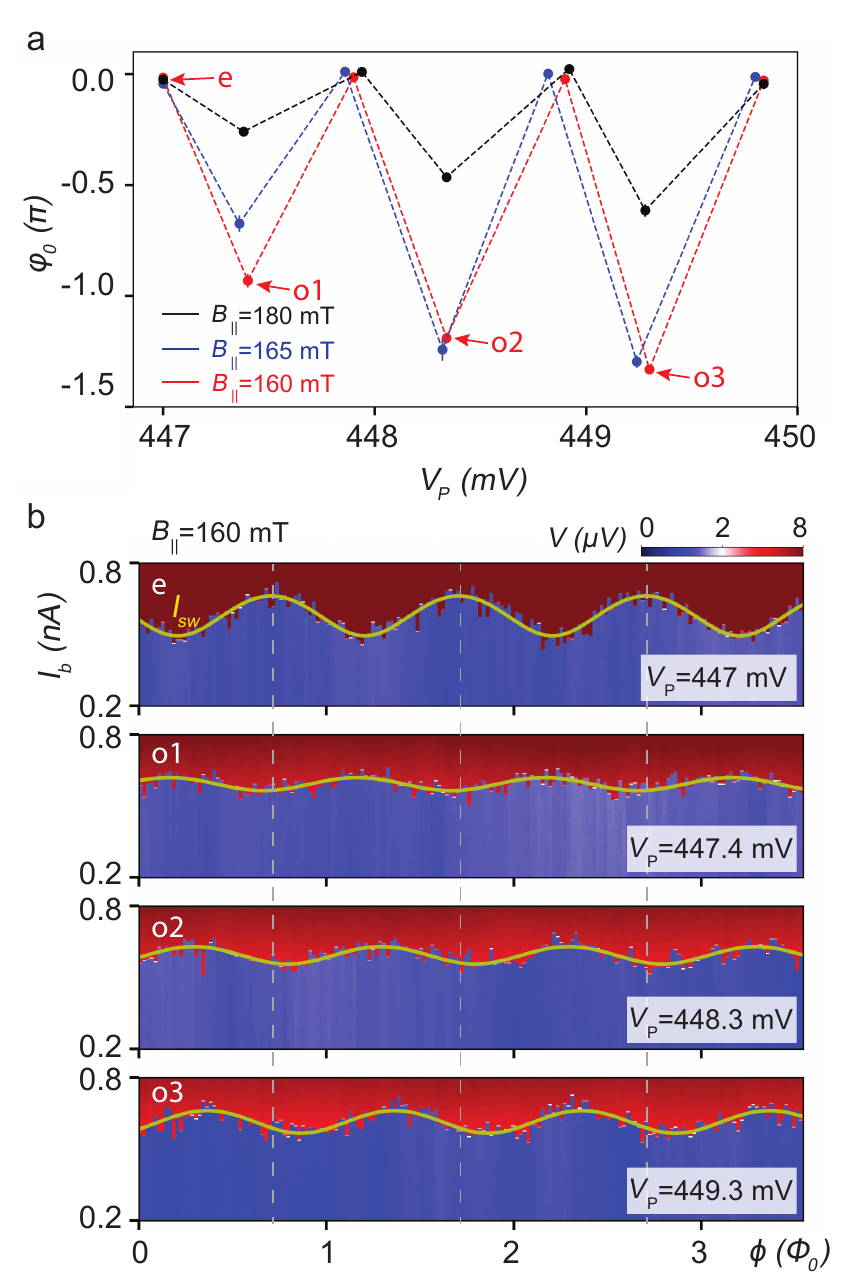}
\caption{(Color online)
\textbf{a}, Phase offset $\varphi_{0}$ versus plunger gate voltage $V_{P}$ for various parallel magnetic fields $B_{\parallel}$. Dashed lines are a guide to the eye. The phase offset is sensitive to both plunger gate voltage and magnetic field variations. 
\textbf{b}, Voltage drop $V$ as a function of the applied bias current  $I_{b}$ and the SQUID flux $\Phi$ for a parallel magnetic field $B_{\parallel}=160\,$mT. The 
switching current $I_{sw}$ (yellow) displays a phase offset $\varphi_{0}$ between even (e) and odd (o) Coulomb valleys of the SC island that is tunable by the plunger gate voltage $V_{P}$. 
}\label{fig:5}
\end{figure}

So far, we have discussed a regime with substantial $\varphi_0$ for parity read-out with maximal resolution. However, such an ideal situation is not always realized. In Fig.~\ref{fig:5}a, we display the phase offset versus plunger gate voltage for multiple magnetic field values. For a selection of data points, we also show the fitted switching current $I_{sw}$ in Fig.~\ref{fig:5}b. Detailed analysis is shown in Fig.S3-Fig.S5 in supplementary material. In comparison, there is another regime in which NW CPT exhibiting phase independence on its parity (see details in Fig.S6-Fig.S7 in supplementary material). In Fig.~\ref{fig:5}, our findings are two-fold: First, we observe that the phase offset for subsequent Coulomb valley pairs is tunable by the magnetic field and the plunger gate voltage. Such a tunability arises because both control parameters change the support of the subgap level wavefunction and, thereby, alter the lead-island Josephson couplings. Second, we find that the phase offset decreases upon increasing the magnetic field. This decrease suggests that the level seperation between the lowest-energy and higher-energy subgap states increases so that the supercurrent contribution with the parity-dependent prefactor becomes energetically unfavorable.  

In summary, we have studied the Josephson relation of an InSb-Al NW CPT. We have demonstrated that upon applying a magnetic field, subgap levels arise in the SC island and mediate a supercurrent with a parity-dependent phase offset. We have shown that the phase offset persists when the subgap state approaches zero energy and enables parity readout of the lowest energy subgap state. Such a supercurrent parity readout could be useful for the manipulation \cite{bib:Beenakker2004,bib:Mao2004,bib:Engel2005,bib:Lalumiere2010,bib:Ionicioiu2007,bib:Pfaff2013} and protection \cite{bib:Andersen2019,bib:Bultink2020} of qubits stored in the isolated subgap levels of SC islands \cite{bib:Aasen2016,bib:Plugge2017,bib:Karzig2017,bib:Schrade20182,bib:Schrade20183}. 

\section{Method}
\noindent\textbf{Device fabrication.} 
The InSb NWs used in the experiment were grown on an Indium phosphide substrate by metalorganic vapor phase epitaxy. In the molecular beam epitaxy chamber, Al flux was deposited along a specific direction to form Al shadows on InSb NWs by neighboring NWs \cite{bib:Gazibegovich2017}. InSb-Al NWs with shadows were transferred onto a doped Si/SiO$_{x}$ substrate using a nano-manipulator installed inside an SEM. NbTiN superconductor was sputter deposited right after Ar etching dedicated to removing the oxidized layer. Subsequently, 30 nm SiN$_{x}$ was sputter deposited to work as a dielectric layer, and 10/120 nm Ti/Au was used as a top gate.  

\noindent\textbf{Transport measurement.} 
The sample was measured at a base temperature of $\sim$20 mK in an Oxford dry dilution refrigerator equipped with a vector magnet. Differential conductance was measured by applying small AC lock-in excitation superimposed on a DC voltage and then measuring AC and DC current through the device. Typically, low frequency of $\sim$ 27 Hz and AC excitation amplitude of $\sim$10 $\mu$V were used for lock-in measurement. In current bias measurement, current was applied through the device while monitoring voltage drop on device. The direction of the magnetic field was aligned with respect to the InSb-Al island arm by detecting the supercurrent of Cooper-pair transistor while rotating the magnetic field direction.

\noindent\textbf{Data taking.} 
In this article, all data comes from one device.     
  
\section{Data availability} 
The authors declare that all of the raw data together with analysis files are available at https://zenodo.org/record/5075186$\#$.YOQqW$\_$kzaUk.
     
\section{Acknowledgement} 
We are grateful to Roman Lutchyn, Bernard van Heck, Michiel de Moor, Chun-Xiao Liu and Marta Pita Vidal for fruitful discussion. This work has been supported by the Dutch Organization for Scientific Research (NWO), the Foundation for Fundamental Research on Matter (FOM) and Microsoft Corporation Station Q. J.S acknowledges support from Chinese Academy of Sciences under Grant No. XDB33000000, National Science Foundation of China under Grant No. 92065203, and  the Synergic Extreme Condition User Facility. 

\section{Author contribution} 
J.W. and J.S. conceived the experiment. J.W. and J.S. fabricated the device. J.W., J.S. and D.V.D. performed the measurements. C.S. and L.F. did the theory simulations. K.L. helped transfer the nanowires. S.G., G.B., R.L.M.O.H.V., J.S.L., M.P., C.P.D., C.J.P. and E.P.A.M.B. carried out the material growth. J.W., J.S., C.S. and V.L. analyzed the data. C.S., J.S. and J.W. wrote manuscript. J.S. supervised the project.      

\section{Competing Interests} 
The authors declare no competing interests.

\end{document}


\hsize\textwidth\columnwidth\hsize\csname@twocolumnfalse\endcsname

\title{Supplemental Material: Supercurrent parity-meter in a nanowire Cooper-pair transistor}

\affiliation{QuTech and Kavli Institute of NanoScience, Delft University of Technology, 2600 GA Delft, The Netherlands}
\affiliation{Department of Physics, Massachusetts Institute of Technology, 77 Massachusetts Avenue, Cambridge, Massachusetts 02139 USA}
\affiliation{Beijing National Laboratory for Condensed Matter Physics, Institute of Physics, Chinese Academy of Sciences, Beijing 100190, China}
\affiliation{Department of Applied Physics, Eindhoven University of Technology, 5600 MB Eindhoven, The Netherlands}
\affiliation{California NanoSystems Institute, University of California Santa Barbara, Santa Barbara, California 93106, USA}
\affiliation{Current address: Department of Physics and Astronomy, University of Tennessee, Knoxville, Tennessee 37996, USA}
\affiliation{Electrical and Computer Engineering, University of California Santa Barbara, Santa Barbara, California 93106, USA}
\affiliation{Materials Department, University of California Santa Barbara, Santa Barbara, California 93106, USA}
\affiliation{Microsoft Station Q Delft, 2600 GA Delft, The Netherlands}
\affiliation{These authors contributed equally}

\author{Ji-Yin Wang}
\address{QuTech and Kavli Institute of NanoScience, Delft University of Technology, 2600 GA Delft, The Netherlands}
\address{These authors contributed equally}

\author{Constantin Schrade}
\address{Department of Physics, Massachusetts Institute of Technology, 77 Massachusetts Avenue, Cambridge, Massachusetts 02139 USA}
\address{These authors contributed equally}

\author{Vukan Levajac}
\author{David van Driel}
\author{Kongyi Li}
\address{QuTech and Kavli Institute of NanoScience, Delft University of Technology, 2600 GA Delft, The Netherlands}

\author{Sasa Gazibegovic}
\author{Ghada Badawy}
\author{Roy L.M. Op het Veld}
\address{Department of Applied Physics, Eindhoven University of Technology, 5600 MB Eindhoven, The Netherlands}

\author{Joon Sue Lee}
\affiliation{California NanoSystems Institute, University of California Santa Barbara, Santa Barbara, California 93106, USA}
\affiliation{Current address: Department of Physics and Astronomy, University of Tennessee, Knoxville, Tennessee 37996, USA}

\author{Mihir Pendharkar}
\address{Electrical and Computer Engineering, University of California Santa Barbara, Santa Barbara, California 93106, USA}

\author{Connor P. Dempsey}
\address{Electrical and Computer Engineering, University of California Santa Barbara, Santa Barbara, California 93106, USA}

\author{Chris J. Palmstrøm}
\address{California NanoSystems Institute, University of California Santa Barbara, Santa Barbara, California 93106, USA}
\address{Electrical and Computer Engineering, University of California Santa Barbara, Santa Barbara, California 93106, USA}
\address{Materials Department, University of California Santa Barbara, Santa Barbara, California 93106, USA}

\author{Erik P.A.M. Bakkers}
\address{Department of Applied Physics, Eindhoven University of Technology, 5600 MB Eindhoven, The Netherlands}

\author{Liang Fu}
\address{Department of Physics, Massachusetts Institute of Technology, 77 Massachusetts Avenue, Cambridge, Massachusetts 02139 USA}

\author{Leo P. Kouwenhoven}
\address{QuTech and Kavli Institute of NanoScience, Delft University of Technology, 2600 GA Delft, The Netherlands}
\address{Microsoft Station Q Delft, 2600 GA Delft, The Netherlands}

\author{Jie Shen}
\email{shenjie@iphy.ac.cn}
\address{QuTech and Kavli Institute of NanoScience, Delft University of Technology, 2600 GA Delft, The Netherlands}
\address{Beijing National Laboratory for Condensed Matter Physics, Institute of Physics, Chinese Academy of Sciences, Beijing 100190, China}

\date{\today}



\maketitle

\begin{widetext}




\begin{center}
{\it ...}
\end{center}

\section{Section1: Additional measurement data} 
\begin{figure*}[!htb]
\centering
\includegraphics[width=0.8\textwidth]{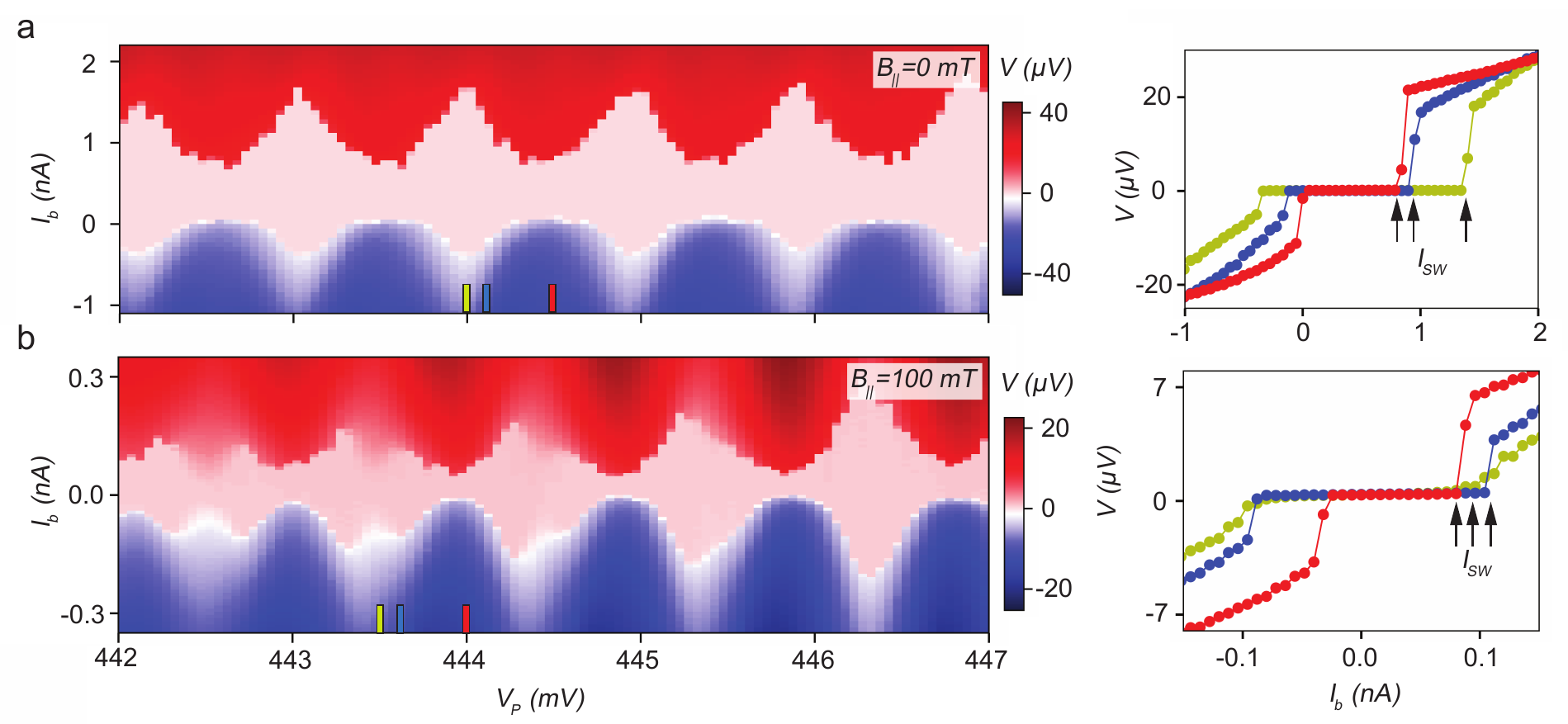}
\caption{(Color online) Current-bias characteristics of the NW CPT at different magnetic fields with the reference arm pinched off. \textbf{a}, Left panel: Voltage drop \textit{V} across the NW CPT as a function of current bias \textit{I$_b$} and plunger gate \textit{V$_P$} at \textit{B$_{||}$}=0. Right panel: Linecuts at three different plunger gate values. \textbf{b}, Left panel: Voltage drop \textit{V} across the NW CPT as a function of \textit{I$_b$} and \textit{V$_P$} at \textit{B$_{||}$}=100 mT. Right panel: Linecuts at three different plunger gate values. Black arrows mark the switching current \textit{I$_{sw}$}, where the NW CPT transitions from the SC state to the normal state.
}\label{fig:S1}
\end{figure*}

\begin{figure*}[!htb]
\centering
\includegraphics[width=0.9\textwidth]{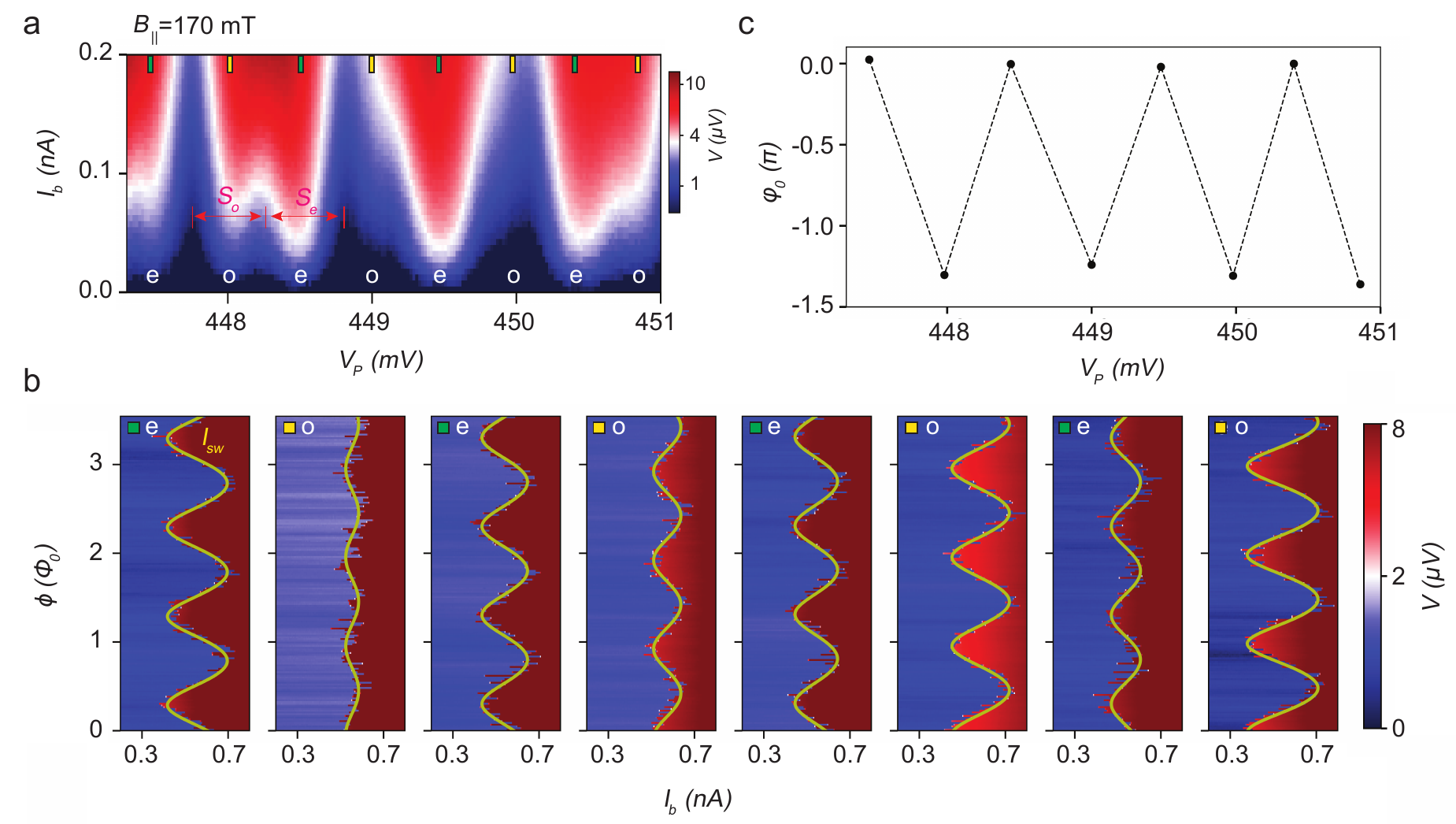}
\caption{(Color online) Transport characteristics of the NW CPT for even and odd charge parity sectors at \textit{B$_{||}$}=170 mT. \textbf{a}, Voltage drop across the NW CPT as a function of current bias \textit{I$_b$} and plunger gate \textit{V$_P$} with reference arm pinched off. Labels `e' (`o') indicate Coulomb valleys of even (odd) charge parities of the SC island. Black bars mark the positions of three most left supercurrent peaks and distance between two neighboring peaks gives peak space for even (\textit{S$_{e}$}) or odd parity (\textit{S$_{o}$}). \textit{S$_{e}$} and \textit{S$_{o}$} have comparable values, indicating lowest sub-gap state is close to zero. \textbf{b}, Voltage drop across the SQUID device as a function of current bias \textit{I$_b$} and flux \textit{$\phi$} threading SQUID loop at different charge parities of the SC island. The fitted switching current (yellow), \textit{I$_{sw}$}, display a phase offset between opposite parity sectors. \textbf{c}, Phase offset \textit{$\varphi_{0}$} versus plunger gate \textit{V$_P$}. Dashed lines are guide lines to eye. The data shown in this figure was measured after a thermal cycle of the dilution refrigerator, while the data in the subsequent figures was measured before the thermal cycle. 
}\label{fig:S2}
\end{figure*}

\begin{figure*}[!htb]
\centering
\includegraphics[width=0.9\textwidth]{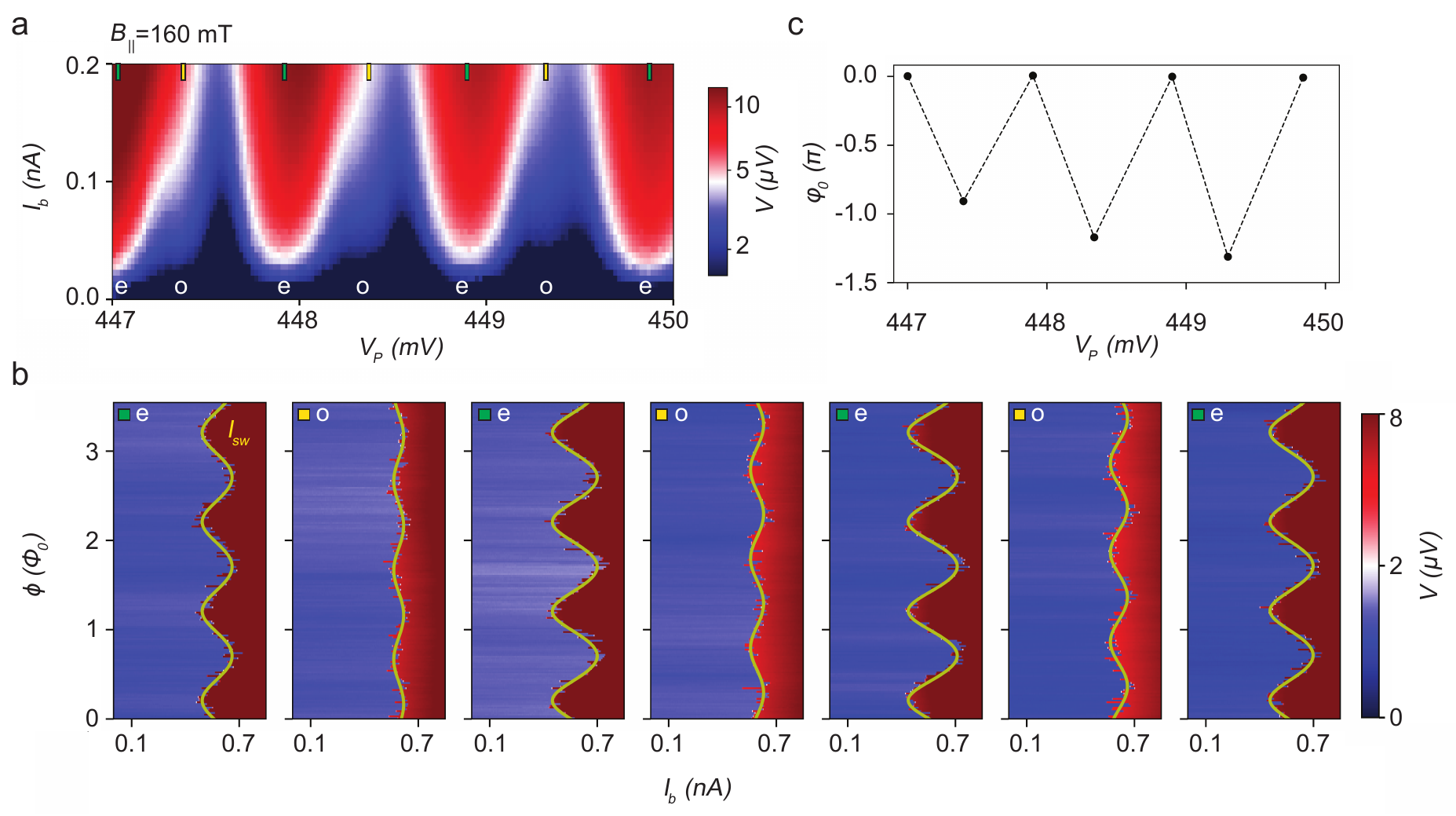}
\caption{(Color online) Transport characteristics of the NW CPT for even and odd charge parity sectors at \textit{B$_{||}$}=160 mT. \textbf{a}, Voltage drop across the NW CPT as a function of current bias \textit{I$_b$} and plunger gate \textit{V$_P$} with reference arm pinched off. Labels `e' (`o') indicate Coulomb valleys of even (odd) charge parities of the SC island. \textbf{b}, Voltage drop across the SQUID device as a function of current bias \textit{I$_b$} and flux \textit{$\phi$} threading SQUID loop at different charge parities of the SC island. The fitted switching current (yellow), \textit{I$_{sw}$}, display a phase offset between opposite parity sectors. \textbf{c}, Phase offset \textit{$\varphi_{0}$} versus plunger gate \textit{V$_P$}. Dashed lines are guide lines to eye. 
}\label{fig:S3}
\end{figure*}

\begin{figure*}[!htb]
\centering
\includegraphics[width=0.9\textwidth]{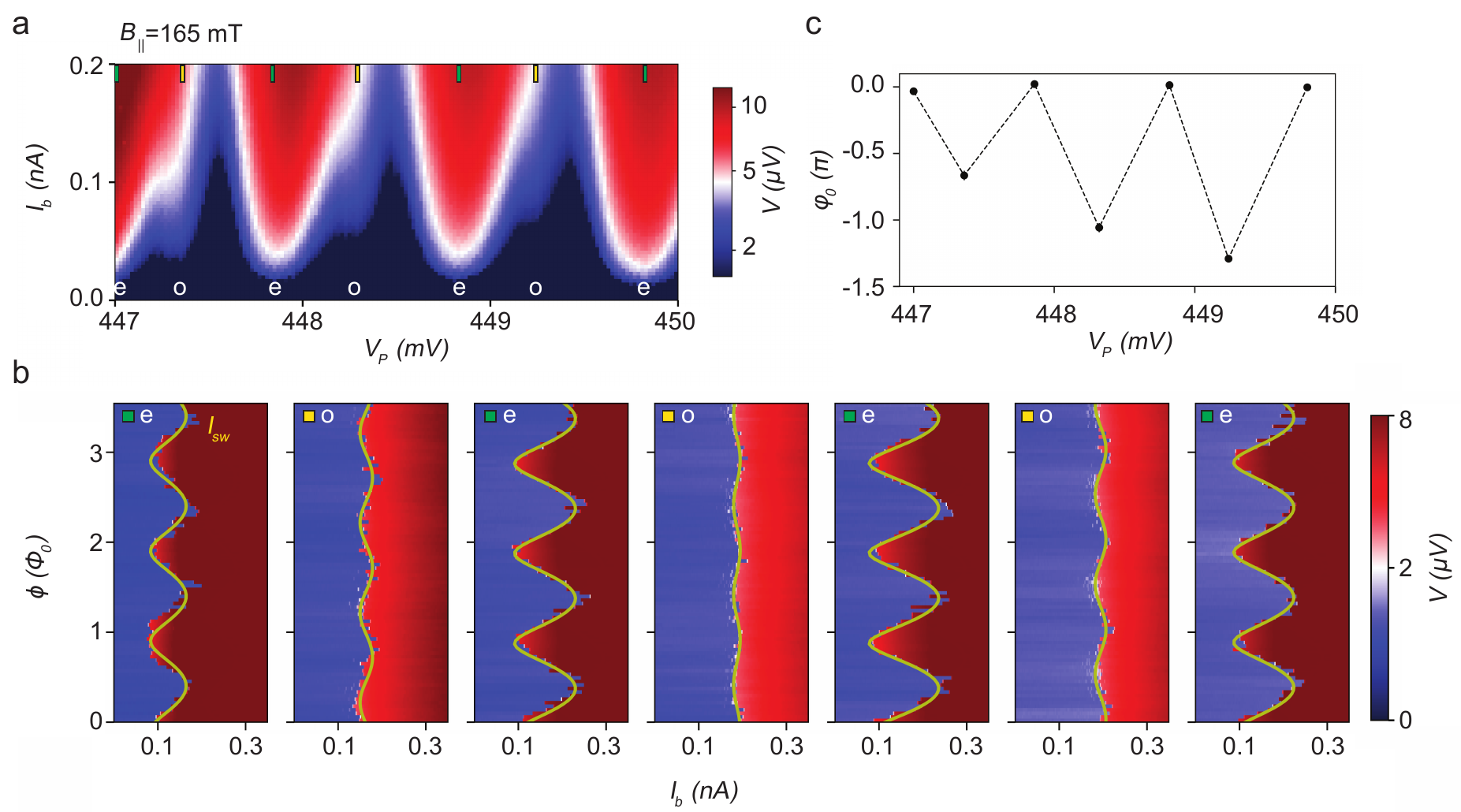}
\caption{(Color online) Transport characteristics of the NW CPT for even and odd charge parity sectors at \textit{B$_{||}$}=165 mT. \textbf{a}, Voltage drop across the NW CPT as a function of current bias \textit{I$_b$} and plunger gate \textit{V$_P$} with reference arm pinched off. Labels `e' (`o') indicate Coulomb valleys of even (odd) charge parities of the SC island. \textbf{b}, Voltage drop across the SQUID device as a function of current bias \textit{I$_b$} and flux \textit{$\phi$} threading SQUID loop at different charge parities of the SC island. The fitted switching current (yellow), \textit{I$_{sw}$}, display a phase offset between opposite parity sectors. \textbf{c}, Phase offset \textit{$\varphi_{0}$} versus plunger gate \textit{V$_P$}. Dashed lines are guide lines to eye. 
}\label{fig:S4}
\end{figure*}

\begin{figure*}[!htb]
\centering
\includegraphics[width=0.9\textwidth]{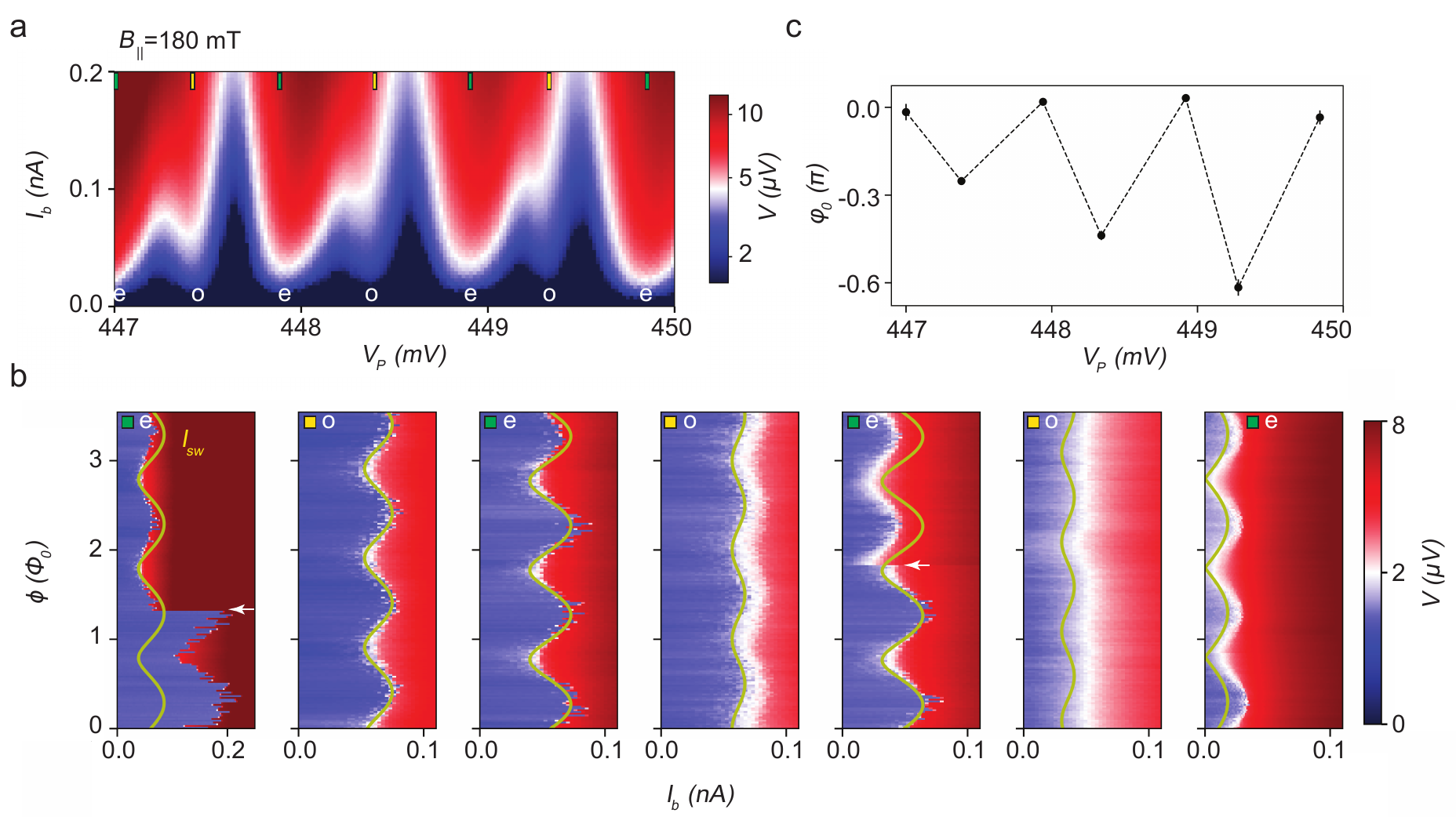}
\caption{(Color online) Transport characteristics of the NW CPT for even and odd charge parity sectors at \textit{B$_{||}$}=180 mT. \textbf{a}, Voltage drop across the NW CPT as a function of current bias \textit{I$_b$} and plunger gate \textit{V$_P$} with reference arm pinched off. Labels `e' (`o') indicate Coulomb valleys of even (odd) charge parities of the SC island. \textbf{b}, Voltage drop across the SQUID device as a function of current bias \textit{I$_b$} and flux \textit{$\phi$} threading SQUID loop at different charge parities of the SC island. The fitted switching current (yellow), \textit{I$_{sw}$}, display a phase offset between opposite parity sectors. White arrows mark where sudden jumps happen, possibly caused by instability of reference gate. In order to reliably fit superconducting phase, the part before or after jump is shifted to align with rest part during fitting. \textbf{c}, Phase offset \textit{$\varphi_{0}$} versus plunger gate \textit{V$_P$}. Dashed lines are guide lines to eye. 
}\label{fig:S5}
\end{figure*}

\begin{figure*}[!htb]
\centering
\includegraphics[width=0.9\textwidth]{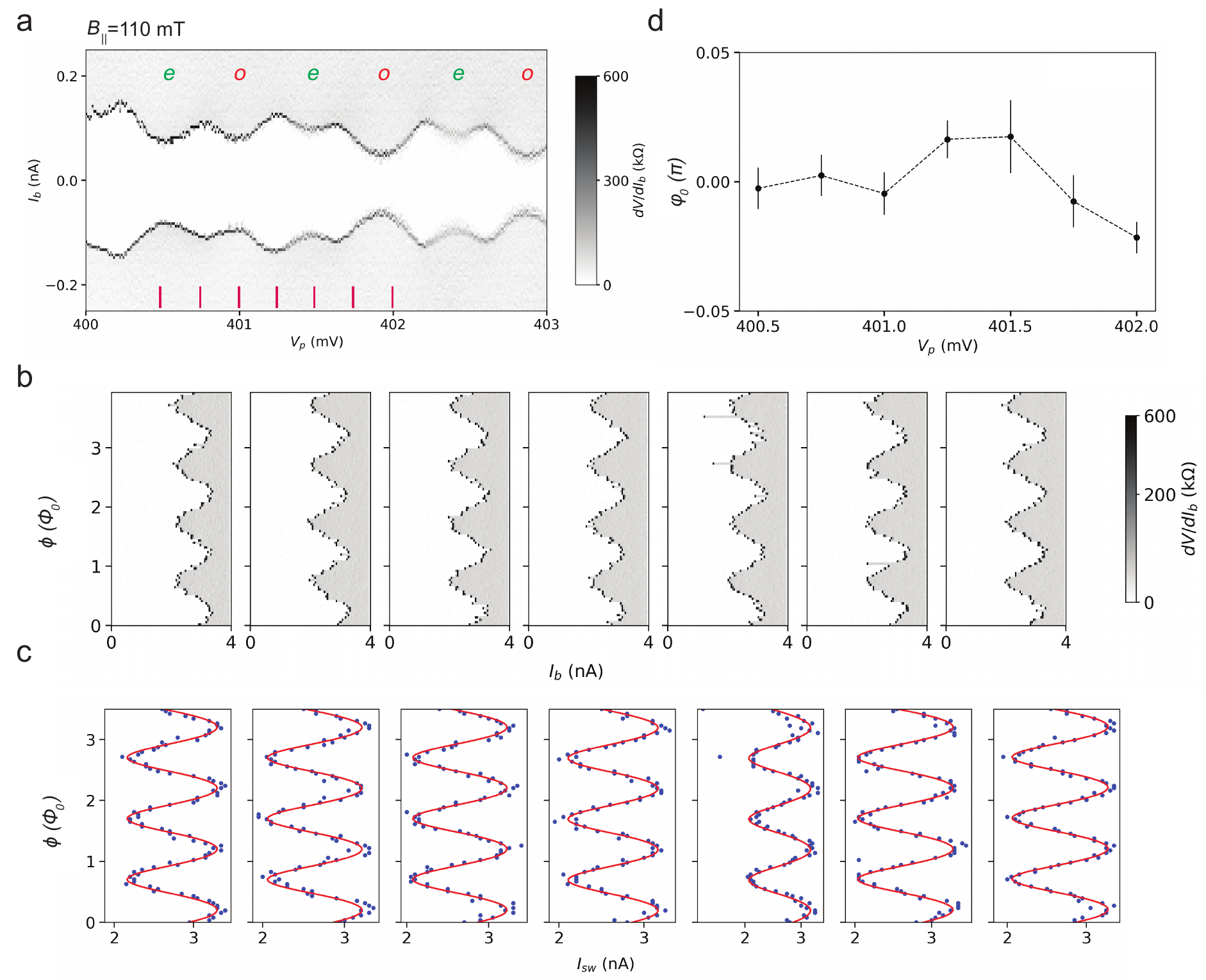}
\caption{(Color online) A different plunger gate \textit{V$_P$} region having phase independence on parities at \textit{B$_{||}$}=110 mT. \textbf{a}, Differential resistance of the NW CPT as a function of current bias \textit{I$_b$} and plunger gate \textit{V$_P$} with reference arm pinched off. Green (red) labels `e' (`o') indicate Coulomb valleys of even (odd) charge parities of the SC island. \textbf{b}, Differential resistance of the SQUID device as a function of current bias \textit{I$_b$} and flux \textit{$\phi$} threading SQUID loop at different plunger gate points marked by red bars in \textbf{a}. \textbf{c}, Extracted switching current (blue points) versus flux \textit{$\phi$} and corresponding fitting curves (red lines). \textbf{d}, Phase offset \textit{$\varphi_{0}$} versus plunger gate \textit{V$_P$}. Dashed lines are guide lines to eye. Note that the values \textit{$\varphi_{0}$} are obtained by subtrating mean value of all points. The varying of phase with \textit{V$_P$} is within error bar fluctuation.  
}\label{fig:S6}
\end{figure*}

\begin{figure*}[!htb]
\centering
\includegraphics[width=0.9\textwidth]{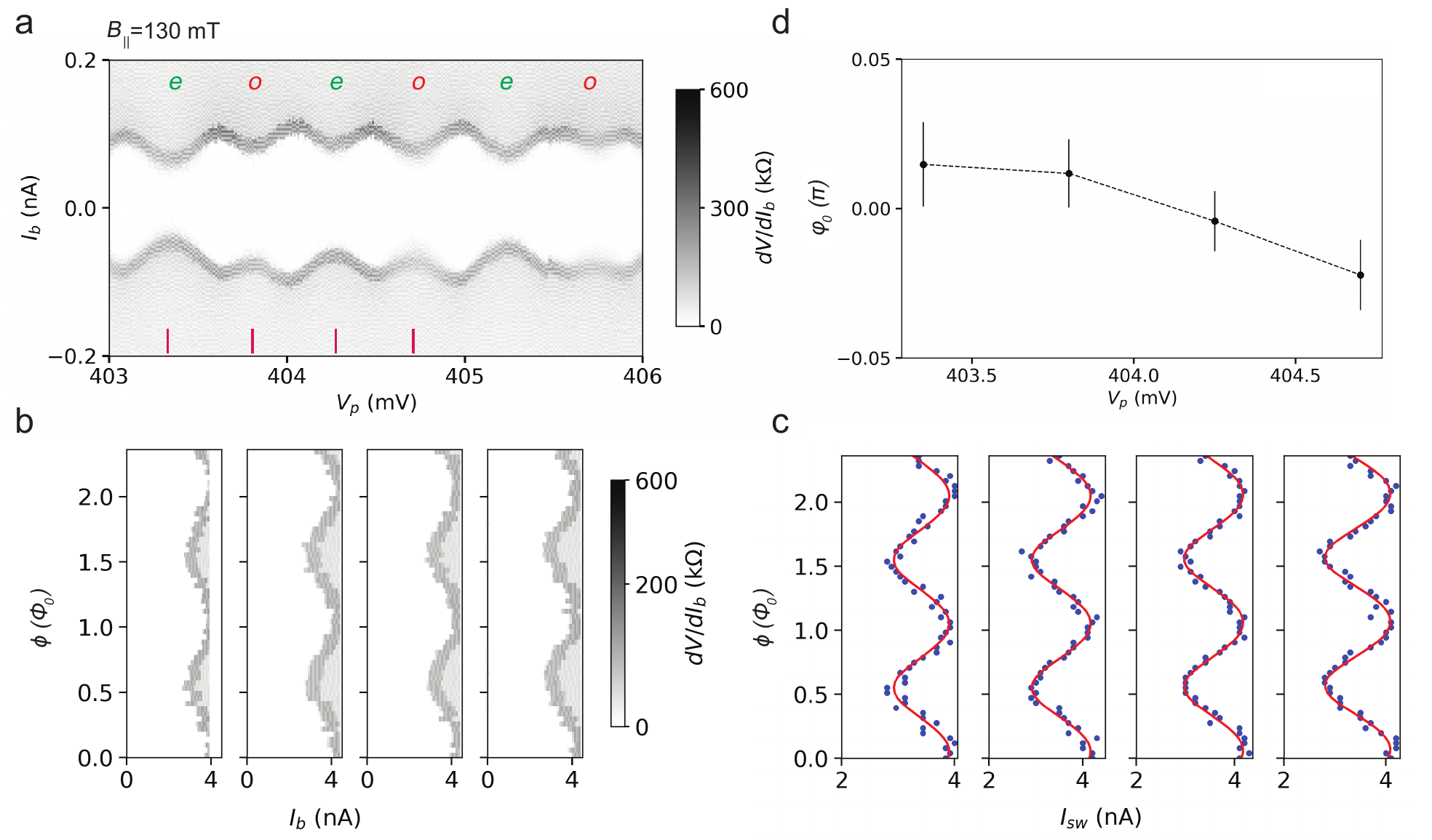}
\caption{(Color online) Another example of phase independence on parities. Plunger gate is very close to that in Fig.~\ref{fig:S6}, and \textit{B$_{||}$}=130 mT. \textbf{a}, Differential resistance of the NW CPT as a function of current bias \textit{I$_b$} and plunger gate \textit{V$_P$} with reference arm pinched off. Green (red) labels `e' (`o') indicate Coulomb valleys of even (odd) charge parities of the SC island. \textbf{b}, Differential resistance of the SQUID device as a function of current bias \textit{I$_b$} and flux \textit{$\phi$} threading SQUID loop at different plunger gate points marked by red bars in \textbf{a}. \textbf{c}, Extracted switching current (blue points) versus flux \textit{$\phi$} and corresponding fitting curves (red lines). \textbf{d}, Phase offset \textit{$\varphi_{0}$} versus plunger gate \textit{V$_P$}. Dashed lines are guide lines to eye. Note that the values \textit{$\varphi_{0}$} are obtained by subtrating mean value of all points. The varying of phase with \textit{V$_P$} is within error bar fluctuation. 
}\label{fig:S7}
\end{figure*}

\FloatBarrier

\section{Section2: Coulomb valley extraction}
We note that the measured current-voltage \textit{I$_{b}$-V} curves of Cooper-pair transistor exhibit a finite slope for all voltages, see Fig.~\ref{fig:S2}a-~\ref{fig:S5}a. The possible reasons are (1) at finite magnetic field, Josephson energy in the Cooper-pair transistor junctions is suppressed and thermal fluctuation results in resistive electron transport \cite{bib:Ambegaokar1969, bib:Jorgensen2007}; (2) In our device, the leads of Cooper-pair transistor are made from NbTiN/Al and NbTiN is able to push quasiparticle into Al, softening Al gap \cite{bib:Drachmann2017}. Coulomb valleys could still be addressed via resistance peak around zero-bias voltage, albiet smeared IV curve resulting from above mentioned two mechanisms, because electron transport is most resistive at Coulomb valleys in both scenarios. Furthermore, both aforementioned mechanisms would not affect superconducting phase measurement results with reference turning on. When reference arm is turned on, total supercurrent (as well as Josephson energy) becomes much larger. Then, thermal fluctuation plays much less of a role and quasi-particle transport is completely suppressed, which is reflected by very sharp transition from superconducting to resistive regime in superconducting phase measurement.  
  
\section{Section3: Discussion on self-inductance of SQUID loop}
According to Tinkham's book \cite{bib:TinkhamBook}, superconducting loop subjected to an external magnetic flux is able to generate screening current to expel external flux out, which could distort the measured current-phase relationship and make precise extraction of superconducting phase difficult. In order to eliminate the doubt, we quantitatively estimate the amplitude of self-generated flux resulting from inductance. In thin superconducting film, total inductance, \textit{L} comprises of kinetic inductance, \textit{L$_{k}$} and geometric inductance, \textit{L$_{g}$}. NbTiN film properties have been systematically studied \cite{bib:JamesThesis}. A typical 100 nm film has Tc of 14 K, resistivity of 123 $\mu\Omega\cdot$cm and \textit{L$_{k}$}/(\textit{L$_{k}$}+\textit{L$_{g}$})$\sim$0.3. In our device, NbTiN has thickness of 80 nm and we adopt the Tc and resistivity from 100 nm film, and \textit{L$_{k}$}/(\textit{L$_{k}$}+\textit{L$_{g}$}) is $\sim$ 0.5 by interpolating the data of \textit{L$_{k}$}/(\textit{L$_{k}$}+\textit{L$_{g}$}) versus film thickness. Kinetic inductance \textit{L$_{k}$} is able to be calulated from eq.(6) in reference \cite{bib:Anthony2010}. To be more critical, we suppose critical current \textit{I$_{c}$} of the loop is 10 nA (actual measured switching current of Cooper-pair transistor is alway bellow 2 nA in our measurement), and the value of \textit{$I_{c}\cdot$L} $\sim$ 4*$10^{-4}\Phi_{0}$, where $\Phi_{0}$ is flux quantum. Thus, self-inductance is negligibly small compared with external flux and was not taken into account during data analysis.

\section{Section4: Effective Hamiltonian for the nanowire Cooper-pair transistor}

In this section of the Supplemental Material, we present more details on the derivation of the effective Hamiltonian for the NW CPT which yields a Josephson relation with parity-dependent phase offset. We proceed in multiple steps:

\subsection{Step 1: Model Hamiltonian}
As a first step, we introduce our model Hamiltonian, which comprises a SC island with subgap states coupled to a pair of $s$-wave SC leads. The Hamiltonian for the SC leads has the form, 
\begin{equation}
H_{\text{SC}}=\sum_{\ell=\text{L,R}}\sum_{\bk} \Psi_{\ell,\bk}^\dagger \left(
\xi_{\bk}\eta_{z}+\Delta_{\ell}\eta_{x}e^{i\varphi_{\ell}\eta_{z}}
\right)\Psi_{\ell,\bk},
\end{equation}
where $\Psi_{\ell,\bk}=(c_{\ell,\bk\ua},c^{\dag}_{\ell,-\bk\da})^{T}$ denotes the Nambu spinor with the electron annihilation operator $c_{\ell,\bk s}$ for momentum $\bk$, spin $s$, and lead $\ell$. Furthermore, $\eta_{x,y,z}$ are the Nambu-space Pauli matrices, and $\xi_{\bk}$ is the normal state dispersion. The magnitudes and the phases of the SC order parameters are $\Delta_{\ell}$ and $\varphi_{\ell}$, respectively. 
For simplicity, we assume that $\Delta_{1}=\Delta_{2}\equiv\Delta$.
\\

Next, we introduce the charging Hamiltonian for the mesoscopic SC island, 
\begin{equation}
 U_{\text{C}}(n) = U\left(n-n_{g}\right)^{2}, 
\end{equation}
where $U$ denotes the charging energy magnitude. Moreover, $n$ is the number operator that counts the electron charges on the island, and $n_{g} $ denotes the induced charge, which is continuously tunable through the plunger gate voltage. As outlined in the main text, we focus on the two lowest energy subgap levels in the SC island, which will mediate the Josephson current between the SC leads. In terms of Majorana operators, $\gamma_{i}=\gamma^{\dagger}_{i}$, the Hamiltonian for the two subgap states reads,
\begin{equation}
H_{\text{SG}}=i\varepsilon_{a}\,\gamma_{1a}\gamma_{2a}+i\varepsilon_{b}\,\gamma_{1b}\gamma_{2b},
\end{equation}
where $\varepsilon_{a,b}$ are the energy splittings. We adjust the induced charge so that the SC island hosts $n_{0}$ electron charges in its ground state and, as a result, the joint fermion parity of the subgap levels satisfies,
\begin{equation}
\gamma_{1a}\gamma_{2a}\gamma_{1b}\gamma_{2b}=(-1)^{n_{0}}.
\end{equation}
\\

Lastly, we introduce the tunneling Hamiltonian to describe the coupling between the SC leads and the SC island,
\begin{align}
H_{\text{T}}
&=
\sum_{\ell,i}
\sum_{\bk s}
\lambda^{s}_{\ell i} \
c^{\dag}_{\ell,\bk s}
\gamma_{i}
e^{-i\phi/2}
+
\text{H.c.}
\end{align}
Here, $\lambda^{s}_{\ell i}$ are complex tunneling amplitudes which connect electrons on the SC lead $\ell$ to the subgap states, which are described by the Majorana operators $\gamma_{i}$. Furthermore, $e^{\pm i\phi/2}$ increases/decreases the number of electrons on the SC island by one unit, $[n,e^{\pm i\phi/2}]=\pm e^{\pm i\phi/2}$, while the Majorana operators $\gamma_{i}$ induce flips of the SC island parity. In summary, the total Hamiltonian for our model is given by, 
 \begin{equation}
H=H_{\text{SC}}+U_{\text{C}}+H_{\text{SG}}+H_{\text{T}}.
\end{equation}

\subsection{Step 2: Effective Hamiltonian (General form)}
As a second step, we provide an overview of the effective Hamiltonian for the model that we introduced in the previous subsection. More specifically, up to fourth order in the tunnel couplings $\lambda^{s}_{\ell i}$, the effective Hamiltonian reads,
\begin{equation}
\begin{split}
H_{\text{eff}}&= 
P\, H_{\text{T}}
\left\{
[
u(n_{a},n_{b})
-
H_{\text{SC}}-U_{\text{C}}-H_{\text{SG}}
]^{-1}
(1-P)
H_{\text{T}}
\right\}^{3}
P.
\end{split}
\end{equation}
Here, $P$ denotes the projection operator on the subspace of $H_{\text{SC}}+H_{\text{C}}+H_{\text{SG}}+H_{\text{T}}$ 
with a fixed charge configuration $(n_{a},n_{b})$ at energy,
\begin{equation}
u(n_{a},n_{b})
=
U_{\text{C}}\left(n_{a}+n_{b}-n_{g}\right)^{2}
+
(-1)^{n_a}\varepsilon_{a}
+
(-1)^{n_b}\varepsilon_{b}.
\end{equation}
To evaluate $H_{\text{eff}}$ based on the equation presented above, we need to compute all sequences of intermediate states that mediate a Cooper pair between the SC leads via the SC island. Before going into the details of this calculation, we first present the general result, 
\begin{equation}
\begin{split}
H_{\text{eff}}&=
-
\gamma_{1a}\gamma_{2a}\gamma_{1b}\gamma_{2b}\left[
\beta
\sum_{m=1}^{4}
J^{(m)}_{ab}
\cos
(
\varphi
+
\varphi^{(m)}_{ab}
)
\right]
-
\alpha_{a}
{J}_{a}
\cos
(
\varphi
+
{\varphi}_{a}
)
-\alpha_{b}
J_{b}
\cos
(
\varphi
+
{\varphi}_{b}
)
\\
&\equiv
-\gamma_{1a}\gamma_{2a}\gamma_{1b}\gamma_{2b}\,\beta\, J_{ab}\cos(\varphi+\varphi_{ab})
-\alpha_{a}J_{a}\cos(\varphi+\varphi_{a})
-\alpha_{b}J_{b}\cos(\varphi+\varphi_{b})
\end{split}
\end{equation}
In the second line, we have defined the Josephson couplings and phase offsets, 
\begin{equation}
\begin{split}
&J_{ab}=\left|\sum_{m=1}^{4}J^{(m)}_{ab}e^{i\varphi^{(m)}_{ab}}\right|
,\qquad
\varphi_{ab}=\text{arg}\left(\sum_{m=1}^{4}J^{(m)}_{ab}e^{i\varphi^{(m)}_{ab}}\right).
\end{split}
\end{equation}
We present expressions for the Josephson couplings $(J_{a},J_{b},J^{(m)}_{ab})$, the phase offsets $(\varphi_{a},\varphi_{b},\varphi^{(m)}_{ab})$, as well as the dimensionless functions $(\alpha_{a},\alpha_{b},\beta)$ in the subsequent sections. Here, we only note that for a fixed charge configuration $(n_{a},n_{b})$, the Josephson relation of the NW CPT is given by,
\begin{equation}
I=
(-1)^{n_0}\,I_{ab}\cos(\varphi+\varphi_{ab})+
I_{a}\cos(\varphi+\varphi_{a})+
I_{b}\cos(\varphi+\varphi_{b}),
\end{equation}
where $I_{ab}=2e\beta\, J_{ab}/\hbar$, $I_{a}=2e\alpha_{a}\, J_{a}/\hbar$, and $I_{b}=2e\alpha_{b}\, J_{b}/\hbar$. This is the result presented in Eq.\,1 of the main text.

\subsection{Step 3: Effective Hamiltonian (Josephson couplings)}
As a third step, we give the expressions for the Josephson couplings $(J_{a},J_{b},J^{(m)}_{ab})$ as well as the dimensionless functions $(\alpha_{a},\alpha_{b},\beta)$, which appear in the previously introduced effective Hamiltonian $H_{\text{eff}}$.
\\

We, initially, define the auxiliary Josephson couplings,
\begin{equation}
J^{i,j}_{k,\ell}=
-\frac{8}{\pi^{2}\Delta}
\left(\sqrt{
\Gamma^{\da}_{\text{L}i}
\Gamma^{\ua}_{\text{L}j}
}
-
\sqrt{
\Gamma^{\da}_{\text{L}j}
\Gamma^{\ua}_{\text{L}i}
}\right)
\left(\sqrt{
\Gamma^{\da}_{\text{R}k}
\Gamma^{\ua}_{\text{R}\ell}
}
-
\sqrt{
\Gamma^{\da}_{\text{R}\ell}
\Gamma^{\ua}_{\text{R}k}
}\right),
\end{equation}
together with the linewidths, 
\begin{equation}
\begin{split}
\Gamma^{s}_{m,1a}&=\pi\nu_{F}|\lambda^{s}_{m,1a}|^{2}\quad,\quad \Gamma^{s}_{m,2a}=\pi\nu_{F}|\lambda^{s}_{m,2a}|^{2} \\
\Gamma^{s}_{m,1b}&=\pi\nu_{F}|\lambda^{s}_{m,1b}|^{2}\quad,\quad \Gamma^{s}_{m,2b}=\pi\nu_{F}|\lambda^{s}_{m,2b}|^{2}.
\end{split}
\end{equation}
Here, for example, $\Gamma^{s}_{m,1a}$ denotes the linewidth that the level $a$ acquires due to the tunneling of electrons with spin $s$ from lead $m$ into $\gamma_{1a}$. The normal-state density of states at the Fermi level for the leads is given by $\nu_{F}$. If, we assume an approximate linewidth $\Gamma^{\text{approx}}=0.01\,$meV and a SC gap $\Delta=0.3\,$meV in the SC leads,
we find an upper-bound estimate for the critical current $I^{\text{approx}}_{c}=[16e(\Gamma^{\text{approx}})^{2}]/(\hbar\pi^{2}\Delta)\approx0.1\,$nA, which is consistent with the values measured in the experiment.
\\

With these definitions, the Josephson couplings, appearing in $H_{\text{eff}}$, are given by,
\begin{equation}
\begin{split}
J_{a}&=J^{1a,2a}_{1a,2a}\quad\quad\ J_{b}=J^{1b,2b}_{1b,2b}\\
J^{(1)}_{ab}&=J^{1a,1b}_{2a,2b}\quad,\quad J^{(2)}_{ab}=J^{1b,2a}_{1a,2b}\quad,\quad J^{(3)}_{ab}=J^{1a,2b}_{1b,2a}\quad,\quad J^{(4)}_{ab}=J^{2a,2b}_{1a,1b}.
\end{split}
\end{equation}

For introducing the dimensionless functions $(\alpha_{a},\alpha_{b},\beta)$, we first define, 
\begin{equation}
g(x)=\sqrt{1+x^2}>0\quad,\quad h(n_{a}+N_{a},n_{b}+N_{b})=\frac{u(n_{a}+N_{a},n_{b}+N_{b})-u(n_{a},n_{b})}{\Delta}>0,
\end{equation}
which allows us to write,
\begin{equation}
\begin{split}
\alpha_{a}
&=
\int_{1}^{\infty}\mathrm{d}x\ \int_{1}^{\infty}\mathrm{d}y\
\frac{
1
}
{
g(x)g(y)
[
g(x)
+
h(n_{a}-1,n_{b})
]
[
g(x)
+
g(y)
]
[
g(x)
+
h(n_{a}+1,n_{b})
]
},
\\
\alpha_{b}
&=
\int_{1}^{\infty}\mathrm{d}x\ \int_{1}^{\infty}\mathrm{d}y\
\frac{
1
}
{
g(x)g(y)
[
g(x)
+
h(n_{a},n_{b}-1)
]
[
g(x)
+
g(y)
]
[
g(x)
+
h(n_{a},n_{b}+1)
]
},
\\
\beta_{1}
&=\frac{1}{4}
\int_{1}^{\infty}\mathrm{d}x\ \int_{1}^{\infty}\mathrm{d}y\
\frac{
1
}
{g(x)g(y)
[
g(y)
+
h(n_{a},n_{b}+1)
]
[
g(x)
+
g(y)
]
[
g(x)
+
h(n_{a}+1,n_{b})
]
}
\\
\beta_{2}
&=\frac{1}{4}
\int_{1}^{\infty}\mathrm{d}x\ \int_{1}^{\infty}\mathrm{d}y\
\frac{1
}
{g(x)g(y)
[
g(x)
+
h(n_{a},n_{b}-1)
]
[
g(x)
+
g(y)
]
[
g(x)
+
h(n_{a}+1,n_{b})
]
}
\\
\beta_{3}
&=\frac{1}{4}
\int_{1}^{\infty}\mathrm{d}x\ \int_{1}^{\infty}\mathrm{d}y\
\frac{
1
}
{g(x)g(y)
[
g(y)
+
h(n_{a}+1,n_{b})
]
[
g(x)
+
g(y)
+
h(n_{a}+1,n_{b}-1)
]
[
g(x)
+
h(n_{a}+1,n_{b})
]
}
\\
\beta_{4}
&=\frac{1}{4}
\int_{1}^{\infty}\mathrm{d}x\ \int_{1}^{\infty}\mathrm{d}y\
\frac{
1
}
{g(x)g(y)
[
g(x)
+
h(n_{a},n_{b}-1)
]
[
g(x)
+
g(y)
+
h(n_{a}+1,n_{b}-1)
]
[
g(x)
+
h(n_{a}+1,n_{b})
]
}
\\
\beta_{5}
&=\frac{1}{4}
\int_{1}^{\infty}\mathrm{d}x\ \int_{1}^{\infty}\mathrm{d}y\
\frac{
1
}
{g(x)g(y)
[
g(x)
+
h(n_{a}-1,n_{b})
]
[
g(x)
+
g(y)
+
h(n_{a}-1,n_{b}+1)
]
[
g(y)
+
h(n_{a}-1,n_{b})
]
}
\\
\beta_{6}
&=\frac{1}{4}
\int_{1}^{\infty}\mathrm{d}x\ \int_{1}^{\infty}\mathrm{d}y\
\frac{
1
}
{g(x)g(y)
[
g(y)
+
h(n_{a},n_{b}+1)
]
[
g(x)
+
g(y)
+
h(n_{a}-1,n_{b}+1)
]
[
g(y)
+
h(n_{a}-1,n_{b})
]
}
\\
\beta_{7}
&=\frac{1}{4}
\int_{1}^{\infty}\mathrm{d}x\ \int_{1}^{\infty}\mathrm{d}y\
\frac{
1
}
{g(x)g(y)
[
g(x)
+
h(n_{a},n_{b}-1)
]
[
g(x)
+
g(y)
]
[
g(y)
+
h(n_{a}-1,n_{b})
]
}
\\
\beta_{8}
&=\frac{1}{4}
\int_{1}^{\infty}\mathrm{d}x\ \int_{1}^{\infty}\mathrm{d}y\
\frac{
1
}
{g(x)g(y)
[
g(y)
+
h(n_{a},n_{b}+1)
]
[
g(x)
+
g(y)
]
[
g(y)
+
h(n_{a}-1,n_{b})
]
}
\\
\beta
&=
\sum^{16}_{p=1}\beta_{p}
.
\end{split}
\end{equation}
In the definition of $\beta$, we have included parameters $\beta_{9},\dots,\beta_{16}$ which are identical to $\beta_{1},\dots,\beta_{8}$ but with the arguments of $h$ interchanged, $h(n,m)\rightarrow h(m,n)$. We note that for a substantial charging energy, virtual states with two additional electrons on the SC island are energetically unfavorable and, for simplicity, have not been accounted for in the expressions for $(\alpha_{a},\alpha_{b},\beta)$.

\subsection{Step 4: Effective Hamiltonian (Anomalous phase shifts)}
As a fourth step, we give the expressions for the phase offsets $(\varphi_{a},\varphi_{b},\varphi^{(m)}_{ab})$ appearing in the effective Hamiltonian $H_{\text{eff}}$.
\\

We, therefore, introduce the auxiliary phase offsets,
\begin{equation}
\varphi^{i,j}_{k,\ell}=\text{arg}[(
\lambda^{\da}_{\text{L}i}\lambda^{\ua}_{\text{L}j}
-
\lambda^{\ua}_{\text{L}i}\lambda^{\da}_{\text{L}j}
)^{*}
(
\lambda^{\da}_{\text{R}k}\lambda^{\ua}_{\text{R}\ell}
-
\lambda^{\ua}_{\text{R}k}\lambda^{\da}_{\text{R}\ell}
)
],
\end{equation}
which allow us express the phase offsets appearing in $H_{\text{eff}}$ as,
\begin{equation}
\begin{split}
\varphi_{a}&=\varphi^{1a,2a}_{1a,2a}\quad,\quad \varphi_{b}=\varphi^{1b,2b}_{1b,2b}\\
\varphi^{(1)}_{ab}&=\varphi^{1a,1b}_{2a,2b}\quad,\quad \varphi^{(2)}_{ab}=\varphi^{1b,2a}_{1a,2b}\quad,\quad \varphi^{(3)}_{ab}=\varphi^{1a,2b}_{1b,2a}\quad,\quad \varphi^{(4)}_{ab}=\varphi^{2a,2b}_{1a,1b}.
\end{split}
\end{equation}

\subsection{Step 5: Effective Hamiltonian (Example calculation)}
As a final step, we provide specific examples on sequences of intermediate states that mediate a contribution to the Josephson current with and without a parity-dependent prefactor. We, thereby, focus on the sequences which we have shown in Fig.~4 of the main text. 
\\

We begin by considering sequences of the type shown in Fig.~4a, which comprise a parity-dependent prefactor. An example, for such a sequence is given by,
\begin{equation}
\begin{split}
&\quad\ P
(
c^{\dag}_{\text{R},-\bq \da}
\gamma_{2,b}
e^{-i\phi/2}
)
(
\gamma_{1,b}
c_{\text{L},-\bk \da}
e^{i\phi/2}
)
(
\gamma_{1,a}
c_{\text{L},\bk \ua}
e^{i\phi/2}
)
(
c^{\dag}_{\text{R},\bq \ua}
\gamma_{2,a}
e^{-i\phi/2}
)
P
\\
&=
P
(
c^{\dag}_{\text{R},-\bq \da}
\gamma_{2,b}
\gamma_{1,b}
c_{\text{L},-\bk \da}
\gamma_{1,a}
c_{\text{L},\bk \ua}
c^{\dag}_{\text{R},\bq \ua}
\gamma_{2,a}
)
P
\\
&=
-
P
(
\gamma_{1,a}
\gamma_{2,a}
\gamma_{1,b}
\gamma_{2,b}
)
(
c^{\dag}_{\text{R},-\bq \da}
c_{\text{L},-\bk \da}
c_{\text{L},\bk \ua}
c^{\dag}_{\text{R},\bq \ua}
)
P
\\
&=
e^{i(\varphi_{\text{L}}-\varphi_{\text{R}})}\,
u_{\bq}v_{\bq}u_{\bk}v_{\bk}\,
P
(
\gamma_{1,a}
\gamma_{2,a}
\gamma_{1,b}
\gamma_{2,b}
)
(
\gamma_{\text{R},\bq \ua}
\gamma_{\text{L},-\bk \da}
\gamma^{\dag}_{\text{L},-\bk \da}
\gamma^{\dag}_{\text{R},\bq \ua}
)
P
\\
&=
e^{i(\varphi_{\text{L}}-\varphi_{\text{R}})}\,
u_{\bq}v_{\bq}u_{\bk}v_{\bk}\,
P
(
\gamma_{1,a}
\gamma_{2,a}
\gamma_{1,b}
\gamma_{2,b}
)
P
\end{split}
\end{equation}
In the third equality, we have represented the electron operators in the SC leads in terms of Bogoliubov quasiparticles through the relations, 
$c_{\ell,\bk \ua}
=e^{i\varphi_{\ell}/2}
(u_{\bk}
\gamma_{\ell,\bk \ua}
+
v_{\bk}
\gamma^{\dag}_{\ell,-\bk \da})$
 and
  $c_{\ell,-\bk \da}=e^{i\varphi_{\ell}/2}
  (u_{\bk}
  \gamma_{\ell,-\bk \da}
  -
  v_{\bk}
  \gamma^{\dag}_{\ell,\bk \ua})$ with the coherence factors $u_{\bk}$, $v_{\bk}$. If we sum over all momenta, we find that the amplitude for the example sequence is given by,
  \begin{equation}
 - (
  \lambda^{\ua}_{\text{L},1a} \lambda^{\da}_{\text{L},1b}
  )^{*}
   ( 
  \lambda^{\ua}_{\text{R},1a} \lambda^{\da}_{\text{R},1b}
  )
  \sum_{\bk,\bq}
\frac{
v_{\bq}
u_{\bk}
u_{\bq}
v_{\bk}
}
{
[E_{\bq}+u(n_{a},n_{b}+1)-u(n_{a},n_{b})][E_{\bk}+E_{\bq}][E_{\bq}+u(n_{a}-1,n_{b})-u(n_{a},n_{b})]
},
  \end{equation}
  where the $E_{\bk}=\sqrt{\xi^{2}_{\bk}+\Delta^{2}}$ denotes the dispersion of the SC leads. If we assume a constant density of states $\nu_{F}$ at the Fermi level, we can rewrite this amplitude as,
  \begin{equation}
 - \frac{1}{\Delta}
\int_{1}^{\infty}\mathrm{d}x\ \int_{1}^{\infty}\mathrm{d}y\
\frac{\nu^{2}_{F}
(
  \lambda^{\ua}_{\text{L},1a} \lambda^{\da}_{\text{L},1b}
)^{*}
( 
  \lambda^{\ua}_{\text{R},1a} \lambda^{\da}_{\text{R},1b}
)
}
{g(x)g(y)
[
g(y)
+
h(n_{a},n_{b}+1)
]
[
g(x)
+
g(y)
]
[
g(y)
+
h(n_{a}-1,n_{b})
]
}.
  \end{equation}
Hence, we conclude that the sequence contributes to the term $\propto\beta_{8}$ in the Josephson relation of the NW CPT.
   \\

Next, we consider sequences of the type shown in Fig.~4b, which do not comprise a parity-dependent prefactor. An example, for such a sequence is given by,
\begin{equation}
\begin{split}
&\quad\ 
P
(
\gamma_{1,a}
c_{\text{L},-\bk \da}
e^{i\phi/2}
)
(
c^{\dag}_{\text{R},-\bq \da}
\gamma_{2,a}
e^{-i\phi/2}
)
(
c^{\dag}_{\text{R},\bq \ua}
\gamma_{2,a}
e^{-i\phi/2}
)
(
\gamma_{1,a}
c_{\text{L},\bk \ua}
e^{i\phi/2}
)
P
\\
&=
P
(
\gamma_{1,a}
c_{\text{L},-\bk \da}
c^{\dag}_{\text{R},-\bq \da}
\gamma_{2,a}
c^{\dag}_{\text{R},\bq \ua}
\gamma_{2,a}
\gamma_{1,a}
c_{\text{L},\bk \ua}
)
P
\\
&=
P
(
c_{\text{L},-\bk \da}
c^{\dag}_{\text{R},-\bq \da}
c^{\dag}_{\text{R},\bq \ua}
c_{\text{L},\bk \ua}
)
P
\\
&=
-
e^{i(\varphi_{\text{L}}-\varphi_{\text{R}})}\,
u_{\bq}v_{\bq}u_{\bk}v_{\bk}\,
P
(
\gamma_{\text{L},-\bk \da}
\gamma_{\text{R},\bq \ua}
\gamma^{\dag}_{\text{R},\bq \ua}
\gamma^{\dag}_{\text{L},-\bk \da}
)
P
\\
&=
-
e^{i(\varphi_{\text{L}}-\varphi_{\text{R}})}\,
u_{\bq}v_{\bq}u_{\bk}v_{\bk}\,
\end{split}
\end{equation}
 If we again sum over all momenta, we find that the amplitude for the example sequence is given by,
  \begin{equation}
(  
  \lambda^{\ua}_{\text{L},1a} \lambda^{\da}_{\text{L},1a}
)^{*}
(
  \lambda^{\ua}_{\text{R},2a} \lambda^{\da}_{\text{R},2a}
)
  \sum_{\bk,\bq}
\frac{
v_{\bq}
u_{\bk}
u_{\bq}
v_{\bk}
}
{
[E_{\bk}+u(n_{a}-1,n_{b})-u(n_{a},n_{b})][E_{\bk}+E_{\bq}][E_{\bk}+u(n_{a}+1,n_{b})-u(n_{a},n_{b})]
}.
  \end{equation}
  In particular, if we assume constant density of states $\nu_{F}$ at the Fermi level, we can rewrite this amplitude as,
    \begin{equation}
    \frac{1}{\Delta}
\int_{1}^{\infty}\mathrm{d}x\ \int_{1}^{\infty}\mathrm{d}y\
\frac{
\nu_{F}^{2}
(  
  \lambda^{\ua}_{\text{L},1a} \lambda^{\da}_{\text{L},1a}
)^{*}
(
  \lambda^{\ua}_{\text{R},2a} \lambda^{\da}_{\text{R},2a}
)
}
{
g(x)g(y)
[
g(x)
+
h(n_{a}-1,n_{b})
]
[
g(x)
+
g(y)
]
[
g(x)
+
h(n_{a}+1,n_{b})
]
}.
  \end{equation}
We, thus, conclude that the sequence contributes to the term $\propto\alpha_{a}$ in the Josephson relation of the NW CPT.

\end{widetext}